\documentclass[fleqn,usenatbib]{mnras}
\usepackage[caption=false]{subfig}
\usepackage{graphicx}
\usepackage{amsmath}
\usepackage{amssymb}
\usepackage[T1]{fontenc}
\usepackage{ae,aecompl}

\newcommand{\be}{\begin{equation}}
\newcommand{\ee}{\end{equation}}
\title[The Multi-Dimensional Structure of Radiative]{The Multi-Dimensional Structure of Radiative Shocks: Suppressed Thermal X-rays and Relativistic Ion Acceleration}

\author[Steinberg \& Metzger]{Elad Steinberg$^{1}$\thanks{E-mail: es3640@columbia.edu} \& Brian D.~Metzger$^{1}$\thanks{E-mail: bdm2129@columbia.edu}
\\
$^{1}$Department of Physics and Columbia Astrophysics Laboratory, Columbia University, New York, NY 10027, USA.
}

\date{Accepted XXX. Received YYY; in original form ZZZ}

\pubyear{2015}

\begin{document}
\label{firstpage}
\pagerange{\pageref{firstpage}--\pageref{lastpage}}
\maketitle

\begin{abstract}
Radiative shocks, behind which gas cools faster than the dynamical time, play a key role in many astrophysical transients, including classical novae and young supernovae interacting with circumstellar material.  The dense layer behind high Mach number $\mathcal{M} \gg 1$ radiative shocks is susceptible to thin-shell instabilities, creating a ``corrugated" shock interface.  We present two and three-dimensional hydrodynamical simulations of optically-thin radiative shocks to study their thermal radiation and acceleration of non-thermal relativistic ions.  We employ a moving-mesh code and a specialized numerical technique to eliminate artificial heat conduction across grid cells.  The fraction of the shock's luminosity $L_{\rm tot}$ radiated at X-ray temperatures $kT_{\rm sh} \approx (3/16)\mu m_p v_{\rm sh}^{2}$ expected from a one-dimensional analysis is suppressed by a factor $L(>T_{\rm sh}/3)/L_{\rm tot} \approx 4.5/\mathcal{M}^{4/3}$ for $\mathcal{M} \approx 4-36$.  This suppression results in part from weak shocks driven into under-pressured cold filaments by hot shocked gas, which sap thermal energy from the latter faster than it is radiated.  Combining particle-in-cell simulation results for diffusive shock acceleration with the inclination angle distribution across the shock (relative to an upstream magnetic field in the shock plane$-$the expected geometry for transient outflows), we predict the efficiency and energy spectrum of ion acceleration.  Though negligible acceleration is predicted for adiabatic shocks, the corrugated shock front enables local regions to satisfy the quasi-parallel magnetic field geometry required for efficient acceleration, resulting in an average acceleration efficiency of $\epsilon_{\rm nth} \sim 0.005-0.02$ for $\mathcal{M} \approx 12-36$, in agreement with modeling of the gamma-ray nova ASASSN-16ma.
\end{abstract}

\section{Introduction}

Non-relativistic shock waves power thermal and non-thermal radiation from a wide range of astrophysical transient events.  The latter include massive star eruptions or young supernovae (SNe) in which the stellar ejecta collides with a dense pre-explosion outflow from the progenitor star (e.g.~SNe IIn; \citealt{Smith&McCray07,Chevalier&Irwin11,Smith16}) or a relic protostellar disk \citep{Metzger10}; the merger of binary stars, or ``common envelope" events (e.g.~\citealt{Ivanova+13,Macleod+18}), following which the dynamically-ejected envelope collides with mass loss from earlier stages of the merger process \citep{Metzger&Pejcha17,Pejcha+17}; and internal collisions within the outflows from hot white dwarfs during classical nova eruptions \citep{Mukai&Ishida01,Sokoloski+06,Martin&Dubus13,Metzger+14,Martin+17}.  In classical novae, relativistic particles accelerated by such shocks power GeV gamma-ray emission, as discovered by {\it Fermi} LAT in temporal coincidence with the nova optical outburst \citep{Ackermann+14,Cheung+16,Franckowiak+18}.  A common feature of all of these systems is that the density of the shocked gas can be sufficiently high for the radiative cooling time to be much shorter than the radial expansion time; such shocks, in which radiative losses play a crucial role in their stability and dynamics, are referred to as ``radiative" (e.g., \citealt{Raymond79,Drake05}).  

As their name implies, the kinetic luminosity dissipated at radiative shocks is radiated with nearly unity efficiency, making them extremely effective at powering luminous transient emission.  This is particularly true when the shocks occur in an expanding outflow that has already lost most of its initial internal energy to adiabatic expansion, a condition which is closely related to the shocks being strong (Mach number $\mathcal{M} \gg 1$).  The shock-powered thermal energy typically emerges from this environment at visual wavelengths, because the column of gas ahead of the shocks effectively absorbs UV and soft X-ray photons emitted by the hot post-shock medium and re-radiates this energy at lower frequencies, where the opacity is much lower.  Nevertheless, thermal X-rays can sometimes be directly observed at sufficiently high photon energies $\gtrsim 1-10$ keV, above the cut-off set by photoelectric absorption  (e.g.~\citealt{Nymark+06,Metzger+14}).  

Though emphasized less frequently, the high gas densities near radiative shocks can also result in efficient non-thermal emission, by providing a dense field of target photons and ambient ions for relativistic particles accelerated near the shock front to interact with via Inverse Compton and hadronic (pion creation) processes, respectively (e.g.~\citealt{Metzger+15,Martin+17}).  Rapid cooling of the thermal plasma also boosts the energy available to non-thermal particles in the post-shock cooling regions through adiabatic compression \citep{Vurm&Metzger18}.  

The dense material surrounding radiative shocks can therefore provide a ``calorimeter" for measuring both the total shock power (from the thermal visual or X-ray wavelength radiation) and the energy in accelerated relativistic particles (from non-thermal X-rays or gamma-rays).  In this way, simultaneous broad-band monitoring of transients thus enables one to constrain or measure the fraction of the shock power which is placed into a power-law distribution of non-thermal particles, i.e.~on the acceleration efficiency $\epsilon_{\rm nt}$ \citep{Metzger+15}.  The classical nova ASASSN-16ma displayed temporally-correlated optical and gamma-ray light curve behavior, a clear signature that the optical radiation was the product of reprocessed radiative shock emission rather than being directly powered by the white dwarf \citep{Li+17}.  Several considerations favor a hadronic origin for the gamma-ray emission from classical novae \citep{Metzger+15,Metzger+16,Martin+17}.  The observed ratio of the optical and gamma-ray flux in ASASSN-16ma implied an ion acceleration efficiency of $\epsilon_{\rm nt} \approx 0.005$ for this event \citep{Li+17}.  

Depending on ones point of view, the acceleration efficiency of $\sim 1\%$ measured for ASASSN-16ma \citep{Li+17} and other classical novae \citep{Metzger+15} is either surprisingly high or surprisingly low.  Particle-in-cell plasma simulations of high Mach number non-relativistic shocks find values of $\epsilon_{\rm nt} \approx 0.05-0.2$ when the magnetic field upstream of the shock is oriented quasi-parallel to the shock normal.  However, no measurable acceleration is seen ($\epsilon_{\rm nt}\approx 0$) when the magnetic field is in the plane of the shock (e.g.~\citealt{Caprioli&Spitkovsky14}).  The latter case is the expected one in the radially-expanding outflows from a transient explosion, because the magnetic field being advected outwards from the central ejecta or pre-explosion stellar wind will be dominated by its toroidal component on large scales due to flux conservation (e.g.~as in the Parker spiral characterizing the solar wind) and thus would be oriented perpendicular to the velocity of the radially-expanding shock front.

Most models of radiative shocks and their emission are based on one-dimensional steady-state hydrodynamical solutions for the flow properties (for example, the MAPPINGS catalog; \citealt{Dopita&Sutherland96,Allen+08}).  However, radiative shocks are prone to various instabilities, which imprint a rich dynamical and geometrical structure that could play an important role in shaping both their thermal and non-thermal emission.  \citet{Chevalier&Immamura82} performed a linear stability analysis of radiative shocks with power-law cooling function ($\Lambda \propto T^{\alpha}$), finding them to be subject to an oscillatory instability for cooling exponents $\alpha \lesssim 0.4$.  \citet{Gaetz+88} explored thermal instability with realistic ISM cooling.  \citet{Innes92} developed one-dimensional models of the thermal instability, finding secondary shocks driven into underpressured, cool gas and finding that magnetic pressure cushions the instability.  The ``pressure-driven thin shell over-stability" (\citealt{Vishniac83,Bertschinger86,MacLow&Norman93}) occurs when the dense shell of gas behind a radiative shock, which is bounded on its other side by a high pressure, is decelerating, causing growing over-stable non-radial oscillations.

Of particular relevance to the present work, the dense cool layer of gas behind a radiative shock is also subject to a ``non-linear thin-shell instability" (NTSI;  \citealt{Vishniac94}).  This occurs because lateral perturbations of the interaction front cause material to be diverted from convex to concave regions, in such a way that the direct compression from the oppositely directed flows creates multiple elongated regions, imparting the shock front with a corrugated geometrical structure.  Numerical simulations of radiative shocks and the NTSI find a shock interaction front characterized by highly complex regions of dense, cooled gas \citep{Stevens+92,Strickland&Blondin95,Walder&Folini00,Pittard09,Parkin+11,McLeod&Whitworth13,Kee+14}.

One consequence of the ``corrugated" shock front created by the NTSI is a reduction in the thermal X-ray luminosity as compared to the naive expectation from a one-dimensional laminar compression analysis (e.g.~\citealt{Kee+14}).  While a head-on strong shock of velocity $v_{\rm sh}$ heats the colliding gas to a temperature $kT_{\rm sh} \approx (3/16)\mu m_p v_{\rm sh}^{2}$, where $\mu$ is the mean molecular weight, this is reduced by a factor $\propto \cos^{2} \alpha$ for shocks of obliquity $\alpha > 0$.  Furthermore, as we will show here, the hot post shock gas undergoes additional losses by performing PdV work on (driving weak shocks into) under-pressurized cold regions, transferring energy to the cool gas before it can be directly radiated by the hot gas and thus further reducing the average temperature of the post-shock emission.  Accurate predictions for the X-ray efficiency of radiative shocks are of crucial importance, as they affect what inferences are made about the power of the shocks from X-ray observations.

Only a few works have attempted to quantify the X-ray suppression of radiative shocks and its dependence on the upstream parameters.  \citet{Kee+14} performed 2D simulations of colliding flows and dual radiative shocks, with Mach numbers $\mathcal{M} \approx 42$, finding suppression of the X-ray luminosity by a factor of $\approx 1/50$ relative to the expectation for an otherwise identical one-dimensional flow.  \citet{Kee+14} make the ansatz that the reduction in the X-ray luminosity scales with the Mach number as $\propto \mathcal{M}^{-1}$, but to our knowledge they provide no motivation for this specific dependency.  

In addition, few of the previous numerical studies demonstrate convergence of their results with grid resolution.  \citet{Parkin&Pittard10} emphasize that ``numerical conduction", and other effects associated with limited grid resolution, can artificially lower the temperature of the shock-heated regions and thus reduce their X-ray emission relative to the physical case.  A fundamental numerical challenge is to resolve the small cooling length behind the shock, which for high Mach number $\mathcal{M} \gtrsim 10-30$ shocks of interest can be orders of magnitude smaller than the system size.  A related challenge is to avoid the artificial exchange of thermal energy between adjacent hot and cold grid cells with densities that differ by factors up to $\sim \mathcal{M}^{2} \gtrsim 10^{2}-10^{3}$. 

Beyond its effect on the temperature of the thermal radiation, the irregular geometry of the radiative shock front could have an equally-pronounced effect on their ability to accelerate relativistic ions and thus to power non-thermal X-ray and gamma-ray emission.  The lower Mach numbers of the oblique shock fronts created by the NTSI could in principle act to steepen the power-law index of the energy distribution of accelerated relativistic particles, $dN/dE \propto E^{-\beta}$, relative to the expectation of $\beta \simeq 2$ in the $\mathcal{M} \gg 1$ limit (e.g.~\citealt{Drury83}).  Furthermore, as discussed above, the local inclination angle of the shock normal relative to the direction of the upstream magnetic field (itself perpendicular to the inflow velocity) plays a crucial role in whether ion acceleration can occur at all (e.g.~\citealt{Caprioli&Spitkovsky14}).  

Here we present high-resolution numerical simulations of radiative shocks using the moving-mesh hydrodynamical code RICH \citep{Yalinewich+15}.  We introduce for the first time a specialized ``Lagrangian" numerical technique, which eliminates artificial diffusion of thermal energy across cell boundaries, allowing us to obtain greater convergence on this problem than possible in previous work.  Our main goal is to quantify the effects of thermal instabilities and the NTSI in radiative shocks on the temperature distribution of the thermal X-ray emission, and on enhancement of non-thermal particle acceleration due to local shock obliquity, as a function of the Mach number.  

This paper is organized as follows.  In $\S\ref{sec:overview}$ we overview the physical properties of radiative shocks and define the domain of applicability of our simulations in the parameter space of the properties of the shock and upstream medium.  In $\S\ref{sec:methods}$ we outline our numerical code and methods, including the ``Lagrangian" scheme to reduce artificial diffusivity, for which we provide numerical verification tests in Appendix \ref{sec:appendix}.  In $\S\ref{sec:results}$ we describe our results and their application to shock-powered transients, such as ASASSN-16ma.  In $\S\ref{sec:discussion}$ we summarize our conclusions and their implications.

\section{Overview of Radiative Shocks}
\label{sec:overview}

Consider two oppositely directed locally-planar flows of equal density $\rho$ which collide head-on, creating dual shocks of velocity $v_{\rm sh}$.  Initially (before cooling instabilities grow and distort the shock front), the shocks are effectively one-dimensional and heat the gas to a temperature given by the usual jump condition
\be
T_{\rm sh} \approx \frac{3}{16}\frac{\bar{\mu}m_p}{k_{b}} v_{\rm sh}^{2} \approx 1.4\times 10^{7}\,{\rm K}\left(\frac{v_{\rm sh}}{10^{3}\,{\rm km\,s^{-1}}}\right)^{2},
\label{eq:Tsh}
\ee
where $\bar{\mu} = 0.62$ for solar composition and $\bar{\mu} = 0.74$ for that typical of classical novae \citep{Vlasov+16}.

\begin{figure}
\centering
\includegraphics[width=0.9\linewidth]{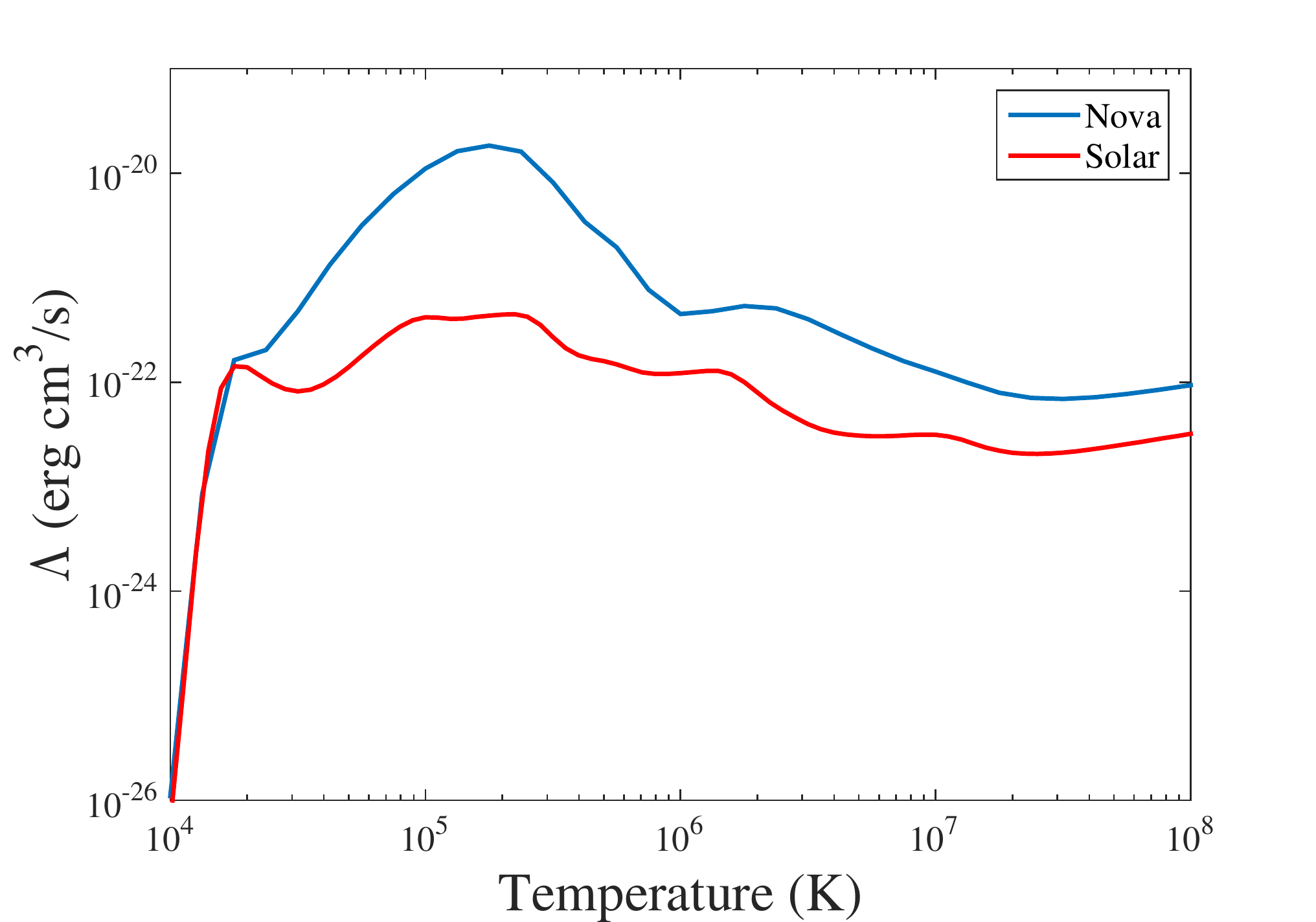}
\caption{Cooling function used in our simulations for solar composition (solid red line) and a composition appropriate to the ejecta of a classical novae (solid blue line), as calculated assuming collisional ionization equilibrium, using CLOUDY version C17.00 \citep{Ferland+17}.  For nova composition, we follow \citet{Vlasov+16} and adopt solar composition with enhanced abundances as follows: [He/H] = 0.08, [N/H] = 1.7, [O/H] = 1.3, [Ne/H] = 1.9, [Mg/H] = 0.7, [Fe/H] = 0.7.}
\label{fig:Lambda}
\vspace{0.4cm}
\end{figure}

We assume that gas cools behind the shock at the optically-thin rate, 
\be \dot{q} = \Lambda(T)n_e n_p,
\label{eq:qdot}
\ee where $n_e$ and $n_p$ are the number densities of free electrons and protons, respectively, and $\Lambda(T)$ is the cooling function (Fig.~\ref{fig:Lambda}).  For a strong shock ($\mathcal{M} \gg 1$) and adiabatic index $\gamma = 5/3$, the gas immediately behind the shock is compressed to a density $\rho_{\rm sh} = 4\rho$ and cools radiatively on the timescale
\be
t_{\rm cool} = \frac{\bar{\mu}}{\mu_p\mu_e}\frac{(3/2)k_{b}T_{\rm sh}}{n_{\rm sh}\Lambda},
\ee
where $n_{\rm sh} = \rho_{\rm sh}/m_p$, $\mu_e = 2m_p/(1+X) \simeq 1.16$ and $\mu_p = 1/X \simeq 1.39 $ for hydrogen mass fraction $X = 0.72$.  The immediate post-shock gas, slowed to a velocity $v_{\rm sh}/4$, cools over a characteristic length-scale given by
\be
\mathcal{L}_{\rm cool} = v_{\rm sh}t_{\rm cool}/4.
\label{eq:Lcool}
\ee
If the interacting flows possess a vertical or radial extent $\sim R$, then define a ``cooling efficiency" according to
\begin{eqnarray}
&\eta& \equiv \mathcal{L}_{\rm cool}/R  \nonumber \\
&\simeq& 10^{-3}\left(\frac{n_{\rm sh}}{10^{10}{\rm cm^{-3}}}\right)^{-1}\left(\frac{v_{\rm sh}}{10^{3}\,{\rm km\,s^{-1}}}\right)^{4}\left(\frac{t_{\rm exp}}{1\,{\rm wk}}\right)^{-1},
\label{eq:eta}
\end{eqnarray}
where we have approximated $\Lambda \approx 1.3\times 10^{-22}(T/10^{7}{\rm K})^{-1}$ erg cm$^{3}$ s$^{-1}$ for nova metallicity gas in the temperature range $10^{5}\,{\rm K} \lesssim T \lesssim 10^{7}$ K.  We normalize the system size to a value $R = v_{\rm sh}t_{\rm exp}$, for a timescale $t_{\rm exp} = 1$ week, and particle density $n_{\rm sh} = 10^{10}$ cm$^{-3}$, values characteristic of Type IIn SNe and classical novae.

A shock is defined as ``radiative" when $\eta \ll 1$  $(t_{\rm cool} \ll t_{\rm exp})$, since in this limit the shocked gas has sufficient time to radiate its thermal energy before adiabatic radial or lateral expansion reconverts the thermal energy back into bulk motion.  In this limit, most of the upstream kinetic power dissipated by the shock,
\be
L_{\rm tot} \sim \bar{\mu} m_p n_{\rm sh} v_{\rm sh}^{3}R^{2},
\ee
ultimately emerges as radiation at some waveband, where the prefactor in this expression depends on the precise geometry of the system.  In the opposite, adiabatic limit $\eta \gg 1$ ($t_{\rm cool} \gg t_{\rm exp}$), the radiative luminosity is reduced from $L_{\rm tot}$ by a factor of $1/(1+5\eta/2)$ for cases when free-free emission dominates the cooling (e.g.~\citealt{Metzger+14}).

If gas behind the shock cools and compresses at roughly constant pressure (as turns out to be a crude but useful approximation), then its density increases by a factor of 
\be
\frac{\rho_{\rm c}}{\rho} \sim \mathcal{M}^{2} \approx \frac{T_{\rm sh}}{T_{\rm min}} \sim 10^{3}\left(\frac{T_{\rm min}}{10^{4}\,{\rm K}}\right)\left(\frac{v_{\rm sh}}{10^{3}\,{\rm km\,s^{-1}}}\right)^{2},
\label{eq:contrast}
\ee

 where $\mathcal{M} = v_{\rm sh}/c_{\rm s}$ is the Mach number of the shock defined relative to the soundspeed $c_{\rm s}$ of the cooled gas.  The latter is generally regulated to a value $\approx (k_{\rm b}T_{\rm min}/\mu m_p)^{1/2} \approx 10-20$ km s$^{-1}$, corresponding to the temperature floor $T_{\rm min} \sim few \times 10^{4}$ K set by photo-ionization balance and the minimum of the cooling curve (Fig.~\ref{fig:Lambda}).  This dense gas, which remains at least partially self-shielded from ionizing radiation, provides an ideal environment for dust nucleation upon further cooling and expansion of the system (e.g.~\citealt{Derdzinski+17}).  Note that equation (\ref{eq:contrast}) neglects pressure support from magnetic fields or non-thermal ions, which may prevent the gas from compressing fully (though even non-thermal particles may cool rapidly, and magnetic fields self-generated at the shock could have a small coherence length and may decay downstream of the shock).

In a one-dimensional model, the thickness of the shell, $w_s$, grows linearly with time $t$ as it collects mass, approximately as $w_s \approx c_{\rm s}t/\mathcal{M}$, and thus reaches a radial thickness $\sim R/\mathcal{M}^{2}$ in a time $t_{\rm exp}$, where $R \sim v_{\rm sh}t_{\rm exp}$ is the approximate radial size of the flow if it is expanding at a velocity $\sim v_{\rm sh}$.

\begin{figure}
\centering
\includegraphics[width=0.9\linewidth]{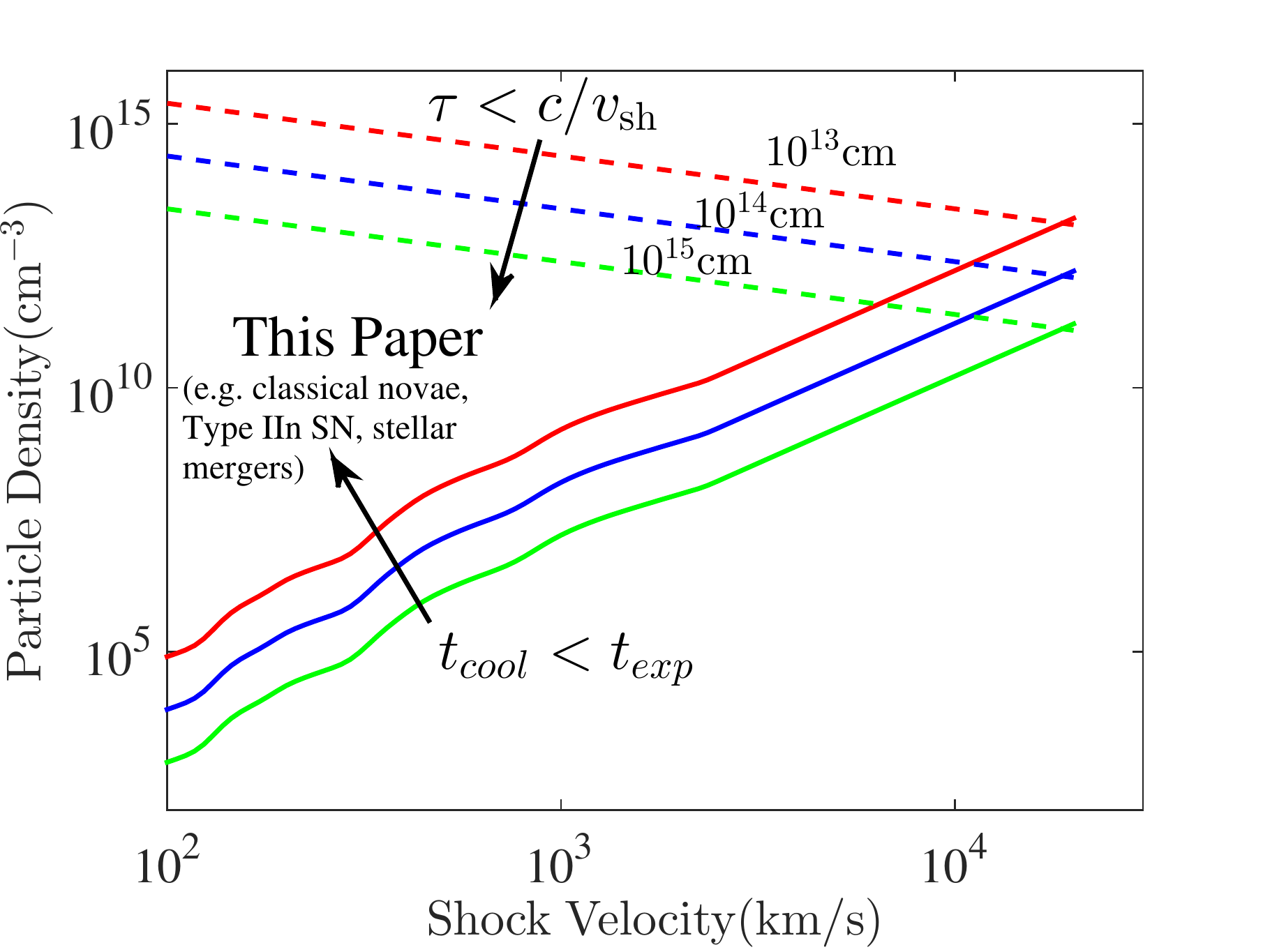}
\caption{The regime of shocks studied in this paper in the space of upstream gas density $n = \rho/(\bar{\mu}m_p)$ and shock velocity $v_{\rm sh}$.  Solid lines show the conditions separating adiabatic from radiative shocks ($t_{\rm cool} = t_{\rm exp}$, where $t_{\rm cool}$ is the optically-thin cooling timescale of the post-shock gas and $t_{\rm exp} = R/v_{\rm sh}$ is the characteristic expansion or evolution timescale) for different values of the shock radius $R = 10^{13}-10^{15}$ cm.  Dashed lines show the condition, $\tau < c/v_{\rm sh}$, for the optically-thin cooling assumption to be valid and for the shock to be mediated by collisionless plasma processes, where $\tau = \rho \kappa R$ is the optical depth to the shock and $\kappa = \kappa_{\rm es} \simeq 0.4$ cm$^{2}$ g$^{-1}$ is the opacity, taken to be electron scattering of fully-ionized hydrogen (this is just a convenient approximation, as the real opacity will likely be the result of Doppler-broadened absorption lines and will be wavelength-dependent).}
\label{fig:schematic}
\end{figure}

Figure \ref{fig:schematic} shows the parameter space of upstream density and shock velocity separating adiabatic from radiative shocks for different ejecta sizes $R = v_{\rm sh}t_{\rm exp}$.  Our implicit assumption that the shocked gas cools at the optically thin rate is a good approximation as long as the timescale for photons to diffuse away from the shock $t_{\rm diff} \sim \tau(R/c)$ is much shorter than the expansion time $t_{\rm exp} \sim R/v_{\rm sh}$, where $\tau \sim \kappa \rho R$ is the optical depth and $\kappa$ is the opacity.  For $t_{\rm diff} \gg t_{\rm exp}$ the energy density of radiation generated near the shock affects its structure (e.g., the shock jump is mediated by photon scattering rather than collisionless plasma processes; e.g.~\citealt{Katz+10}), a case not studied here.  The condition $\tau = c/v_{\rm sh}$ which separates ``effectively transparent" from radiation-mediated shocks are shown by dashed lines in Figure \ref{fig:schematic}.  The triangular wedge between the solid and dashed lines, as characterizes many shock-powered astrophysical transients, represents the region studied in this paper.

In this paper we neglect thermal conduction, which introduces a cooling term to the hot post-shock gas of the form $\nabla \cdot \vec{F}_{c}$ in addition to radiative losses, where $\vec{F}_{c}$ is the conductive flux.  Neglecting magnetic fields, the latter can be written as 
\be
\vec{F}_c = \nabla (\kappa T),
\ee 
where $\kappa = \kappa_0 T^{5/2}$ is the conductivity, $\kappa_0 \approx (2\times 10^{-5}/{\rm ln}\, \Lambda_{\rm c}$) erg K$^{-7/2}$ s$^{-1}$ cm$^{-1}$ is a constant, and ${\rm ln}\,\Lambda_{\rm c} \approx 10$ is the Coulomb logarithm.   Under the assumption that temperature gradients are comparable to the cooling length, i.e.~$\nabla \sim 1/\mathcal{L}_{\rm cool}$, we can approximate $\nabla \cdot \vec{F}_{c} \sim \kappa_0 T^{7/2}/(\mathcal{L}_{\rm cool})^{2}$, where $\mathcal{L}_{\rm cool} \approx (k_b T v)/(n\Lambda) \sim (k_b T^{2} v_{\rm sh})/(n\Lambda T_{\rm sh})$.  Here we have generalized equation (\ref{eq:Lcool}) to account for the expectation, for a one-dimensional flow, that the velocity behind the shock $v = v_{\rm sh}(T/T_{\rm sh})$ decreases due to mass flux conservation $n v = constant$ as gas compresses at roughly constant pressure $nT = constant$.  

The ratio of the conductive to radiative cooling rates is therefore approximately given by
\begin{eqnarray}
\frac{\nabla \cdot \vec{F}_{c}}{n^{2} \Lambda(T)} &\sim& \frac{\kappa_0 T_{\rm sh}^{2}\Lambda}{k_b^{2} T^{1/2} v_{\rm sh}^{2}} \nonumber \\
&\approx& 0.05 \left(\frac{T}{T_{\rm sh}}\right)^{-3/2}\left(\frac{v_{\rm sh}}{10^{3}\,{\rm km\,s^{-1}}}\right)^{-1},
\label{eq:conduction}
\end{eqnarray}
where we have again used the nova composition approximation for $\Lambda$.  Conduction thus provides a moderate perturbation to radiative cooling in the immediate post-shock gas where $T \approx T_{\rm sh}$, but could become more important at lower temperatures $T \ll T_{\rm sh}$.  However, also note that even a weak magnetic field greatly suppresses thermal conduction perpendicular to the local field direction (e.g.~\citealt{Narayan&Medvedev01}).

\section{Numerical Method and Simulation Runs}
\label{sec:methods}

\begin{table*}
\begin{center}
\caption{Simulation Runs}
\begin{tabular}{cccccc}
\hline
\hline

Model & $v_{\rm sh}$ &  $\mathcal{M}^{(a)}$ & Cooling$^{(b)}$ & Cell Size$^{(c)}$ & Num.$^{(d)}$ \\
\hline
\hline
 & (km s$^{-1}$)  & & & ($\mathcal{L}_{\rm cool}$) & \\
\hline
M36\_L & 500 & 36 & Nova & 400 & L \\
M36\textsubscript{Sol}\_L & - & - & Solar & - & - \\
${\dagger}$M36\textsubscript{3D}\_L & - & - & Nova & 300 & - \\
M36\textsubscript{LowRes}\_L & - & - & - & 280 & - \\
M28\_L & 387 & 28 & - & 400 & - \\
M22\_L & 275 & 22 & - & - & - \\
M12\_L & 162 & 12 & - & - & - \\
M4\_L & 50 & 4 & - & 30 & - \\
M36\_SL & 500 & 36 & - & 400 & SL \\
M28\_SL & 387 & 28 & - & - & - \\
M22\_SL & 275 & 22 & - & - & - \\
M12\_SL & 162 & 12 & - & - & - \\
M4\_SL & 50 & 4 & - & 30 & - \\
\hline
\hline
\label{table:runs}
\end{tabular}
\\
$^{(a)}$Mach number; $^{(b)}$Composition assumed in cooling function (Fig.~\ref{fig:Lambda}); $^{(c)}$Initial cell size, in units of the initial cooling length $\mathcal{L}_{\rm cool}$; $^{(d)}$Numerical technique: ``Lagrangian" (L) or ``Semi-Lagrangian" (SL);
$^{\dagger}$3D simulation.
\end{center}
\end{table*}

\begin{table*}
\begin{center}
\caption{Key Results of Simulations}
\begin{tabular}{cccccccc}

\hline
\hline

Model & $T_{\rm sh}^{(a)}$ (K) & $\langle T \rangle^{(b)}$ (K) &  $L(>T_{\rm sh}/3)/L_{\rm tot}^{(c)}$ & $\langle \epsilon_{\rm nth} \rangle_{\theta_{\rm max} = 40^\circ}^{(d)}$ & $\langle \epsilon_{\rm nth} \rangle_{\theta_{\rm max} = 55^\circ}^{(d)}$ & $\langle q-4 \rangle ^{(e)}$ & $L_{\rm shock}/L_{\rm tot}^{(f)}$\\
\hline
M36\_L & $7.2 \times 10^{6}$ K & $3.3\times 10^{5}$ K & $3.6\times10^{-2}$ & $1.5\times10^{-2}$ & $2.7\times10^{-2}$ &$6.5\times10^{-3}$ & 0.82\\
M36\textsubscript{3D}\_L & - & $1.2\times 10^{5}$ K & $9.5\times10^{-3}$ & $10^{-2}$ & $2.2\times10^{-2}$ &$4.3\times10^{-3}$ &0.63 \\
M36\textsubscript{LowRes}\_L & - & $4.4\times 10^{5}$ K & $5.6\times10^{-2}$ & $8\times10^{-3}$ & $1.5\times10^{-2}$ &$7\times10^{-3}$ &0.7 \\

M36\textsubscript{Sol}\_L & - & $3.4\times 10^{5}$ K & $3.3\times10^{-2}$ & $9.6\times10^{-3}$&$1.9\times10^{-2} $& $7.5\times10^{-3}$ &0.77 \\
M28\_L & $4.3 \times 10^{6}$ K & $3\times 10^{5}$ K & $5.9\times10^{-2}$ & $8.8\times10^{-3}$&$1.8\times10^{-2}$ & $10^{-2}$  &0.86\\
M20\_L & $2.2 \times 10^{6}$ K & $1.75\times 10^{5}$ K & $7.38\times10^{-2}$ &$ 1.8\times10^{-3} $&$4.9\times10^{-3}$& $2.2\times10^{-2}$  &0.79\\
M12\_L & $7.5 \times 10^{5}$ K & $1.1\times 10^{5}$ K & $1.6\times10^{-1}$ &$ 4\times10^{-3}$ &$9.2\times10^{-3}$& $8.5\times10^{-2} $ &0.72\\
M4\_L & $7.2 \times 10^{4}$ K & $3.3\times 10^{4}$ K & 0.72 & 0 &0& - & 1 \\

M36\_SL & $7.2 \times 10^{6}$ K & $5\times 10^{4}$ K & $1.2\times10^{-3} $&$ 2.2\times10^{-2} $&$ 3.2\times10^{-2} $ & $8.1\times10^{-3}$  &0.59\\
M28\_SL & $4.3 \times 10^{6}$ K & $4.7\times 10^{4}$ K & $2.9\times10^{-3}$ &$ 2.2\times10^{-2}$&$3.3\times10^{-2}$ & $1.1\times10^{-2}$ &0.69 \\
M20\_SL & $2.2 \times 10^{6}$ K & $6.6\times 10^{4}$ K & $1\times10^{-2}$ &$ 8.8\times10^{-3}$ &$1.6\times10^{-2}$& $2.2\times10^{-2}$  &0.52\\
M12\_SL & $7.5 \times 10^{5}$ K & $1.1\times 10^{5}$ K & $1.5\times10^{-1} $&$ 2.7\times10^{-3}$&$6.7\times10^{-3}$ & $8.6\times10^{-2}$ &0.57 \\
M4\_SL & $7.2 \times 10^{4}$ K & $3.3\times 10^{4}$ K & 0.72 & 0 & 0 & - & 1 \\
\hline
\hline
\label{table:results}
\end{tabular}
\end{center}
$^{(a)}$ Post-shock temperature for the head-on collision of a one-dimensional flow with the simulation parameters (eq.~\ref{eq:Tsh}), as approximately achieved at early times when the shock is still adiabatic; $^{(b)}$ Average temperature of the radiating gas, as weighted by the radiated luminosity, evaluated once the NTSI has reached saturation; $^{(c)}$Fraction of the shock's luminosity $L_{\rm tot}$ emitted at temperatures $> T_{\rm sh}/3$, evaluated once the NTSI has reached saturation; $^{(d)}$Average non-thermal ion acceleration efficiency across the shock front, as weighted by the local shock power, assuming the maximum angle between the shock inclination angle and the upstream magnetic field for particle acceleration is 40$^{\circ}$ or 55$^{\circ}$, respectively (these correspond to the approximate range of cut-off angles found by \citealt{Caprioli&Spitkovsky14}).  This quantity is also averaged over time, after the NTSI has reached saturation; $^{(e)}$Average power-law index of the momentum distribution of accelerated non-thermal ions, as weighted by the local shock power and measured relative to the value $q = 4$ predicted by linear diffusive shock acceleration in the strong shock limit $\mathcal{M} \gg 1$ (eq.~\ref{eq:r}). $^{(f)}$Fraction of the kinetic energy of the upstream flow which is dissipated into thermal energy at the outer shock fronts; the remaining fraction is instead dissipated through weaker shocks or turbulence in the post-shock region.
\end{table*}

We simulate the evolution of radiative shocks using the moving-mesh hydrodynamical code RICH \citep{Yalinewich+15}.  The moving-mesh technique is ideal for this problem because of the large compression factors involved.  RICH solves the compressible Euler equations on a moving Voronoi mesh via a second order Gudonov scheme.  The boundary conditions are an inflowing gas with a velocity of $v_{\rm sh}$ and an adiabatic index of $\gamma = 5/3$.  The density and temperature of the upstream flow are in all cases set to $\rho = 2.6\cdot10^{-12}$ g/cm$^{3}$ and $T = T_{\rm min} = 10^4$ K, respectively.  

Our suite of numerical runs are summarized in Table \ref{table:runs}, labeled according to Mach number and numerical technique (see below).  We cover a range of shock velocities ($v_{\rm sh} \approx 50-500$ km s$^{-1}$), Mach numbers ($\mathcal{M} \sim 4-36$), cooling functions (both solar and classical nova ejecta abundances), and grid resolutions.  Note that once the shocked gas undergoes the cooling instability, the velocity of the shocks decreases by a factor of $3/4$, since the unstable shock front expands at a much lower velocity $v_{\rm sh}/\mathcal{M} \ll v_{\rm sh}$ towards the upstream direction once it has lost pressure support rather than the value $v_{\rm sh}/4$ for an adiabatic shock.  Since the shock starts out adiabatic at early times $t<t_{\rm cool}$, we therefore define $T_{\rm sh}$ to be a factor $16/9$ higher than the one defined in eq. \ref{eq:Tsh} for all of our runs (except the lowest Mach number runs, which are stable).  Also note that the grid resolutions in Table \ref{table:runs} are defined with respect to the initial state of the upstream gas, prior to when the gas has cooled and compressed.  Although our grid cells compress with the flow, we typically achieve a resolution in the compressed regions that is a factor of $\sim$1/$\mathcal{M}$ smaller than the initial upstream one.  Most of our simulations are performed in two spatial dimensions, motivated by the largely two-dimensional nature of the NTSI.  We also perform a single 3D simulation, which, as we shall describe, undergoes a qualitatively similar evolution compared to the otherwise equivalent 2D case.

For reasons described above (Fig.~\ref{fig:schematic}), the shocked gas is assumed to cool at the optically-thin rate (eq.~\ref{eq:qdot}).  While this assumption is not valid for the UV and X-ray emission directly emitted by the hot gas, absorption and reprocessing of the emission occurs very rapidly to visual frequencies, where the opacity and diffusion time is much lower $t_{\rm diff} \ll t$.  The colliding medium is therefore effectively optically thin for purposes of the hydrodynamic evolution.   Figure \ref{fig:Lambda} shows the cooling functions used, calculated from version C17.00 of CLOUDY \citep{Ferland+17} using the \textsc{nova} and \textsc{GASS10} abundances. The cooling is implemented via the exact cooling method proposed in \cite{Townsend}.

Previous works on colliding radiative shocks (e.g.~\citealt{Kee+14}) find a large reduction in the X-ray luminosity arising from radiative shocks \citep{Vishniac94}.  However, some of this reduction may not be physical due to artificial numerical diffusion of thermal energy from hot to cold cells \citep{Parkin&Pittard10}.  There are two ways to overcome this issue. First, one may take the brute force approach of increasing the number of grid cells per cooling length; however, the required resolution depends on the Mach number and can be prohibitively high for $\mathcal{M} \gg 1$. The second method, which we have decided to adopt, is to eliminate the advection term in the numerical scheme altogether.  Though this leads to moderate numerical inaccuracies, we find that it preserves the main qualitative features of the shock evolution while allowing numerical convergence on the reduction of the X-ray emission. We refer to this new approach as the ``Lagrangian" scheme, in contrast to our simulations employing the standard formulation with the advection term included, which we refer to as ``semi-Lagrangian".  Appendix \ref{sec:appendix} provides a detailed description of the ``Lagrangian" scheme and its validation.


To diagnose the structure of the shock and its role in heating the gas, we locate the external shock fronts by identifying cells that have a density within $3\%$ and pressure within $10\%$ of the original inflow parameters, as well as a large pressure gradient and a negative velocity divergence. Once those cells have been identified, we locate the cell with the highest temperature that is within 3 cells distance and along the pressure gradient direction, thus also defining the local shock direction. From the temperature ratio between the upstream and downstream cells we calculate the Mach number of the shock, using the regular jump conditions across shocks.  We have also tried using the procedure outlined by \citet{Schaal+15} and \cite{Schaal+16} to detect shocks; however, this gave many potential spurious detections in the turbulent region between the outer forward and reverse shocks.

\section{Results}
\label{sec:results}
Our results are summarized in Figures \ref{fig:walkthrough}-\ref{fig:spectrum} and Table \ref{table:results}.  We begin with a walk-through of the time evolution of the shock interaction and development of the NTSI for our fiducial run employing the ``Lagrangian" technique with $\mathcal{M} = 36$ ($\S\ref{sec:walkthrough}$).  We then discuss the implications of our suite of results for the temperature distribution of the shock luminosity ($\S\ref{sec:temperature}$) and the acceleration of non-thermal ions ($\S\ref{sec:acceleration}$).

\subsection{Time Evolution and Saturated State}
\label{sec:walkthrough}
\begin{figure}
\centering
\includegraphics[width=0.9\linewidth]{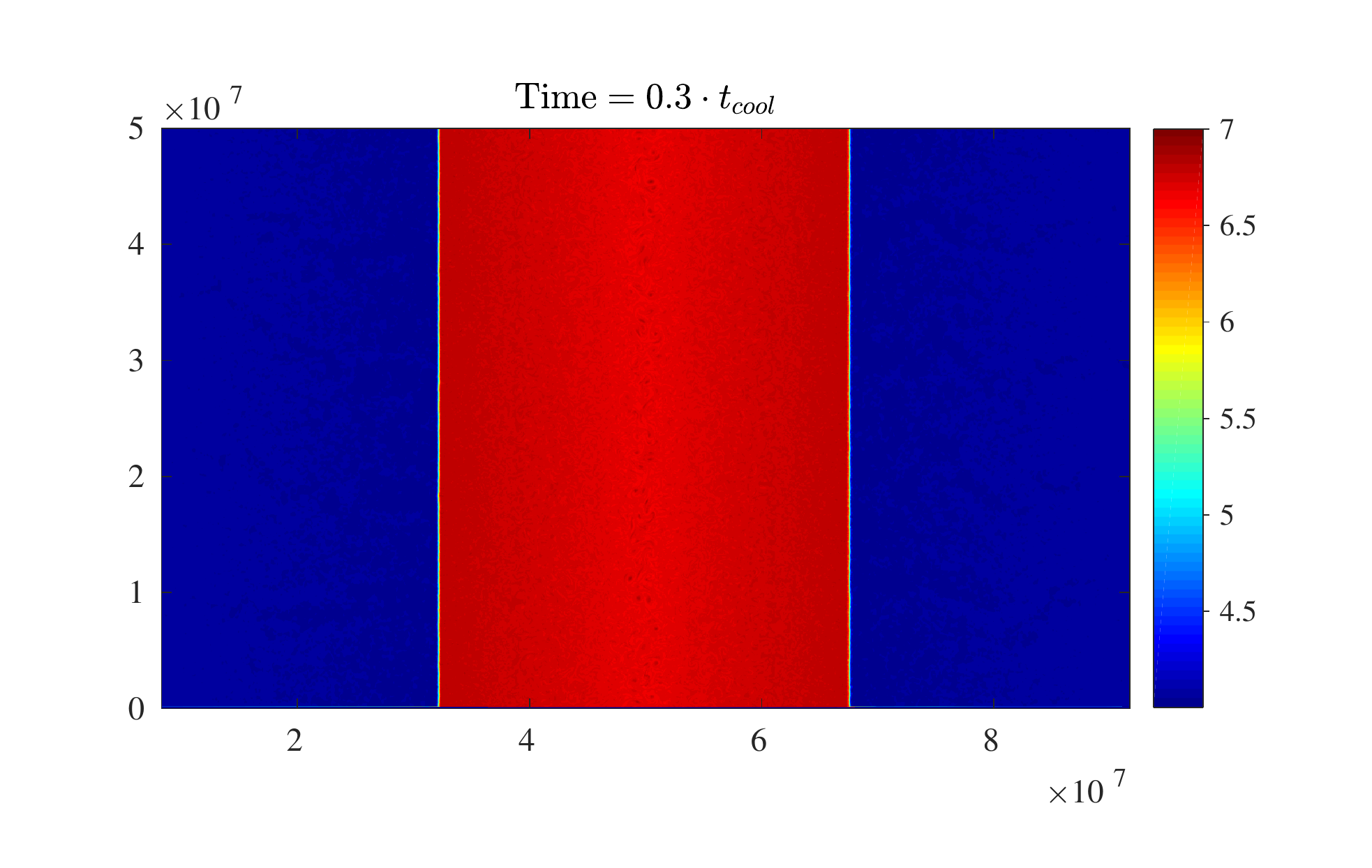}

\includegraphics[width=0.9\linewidth]{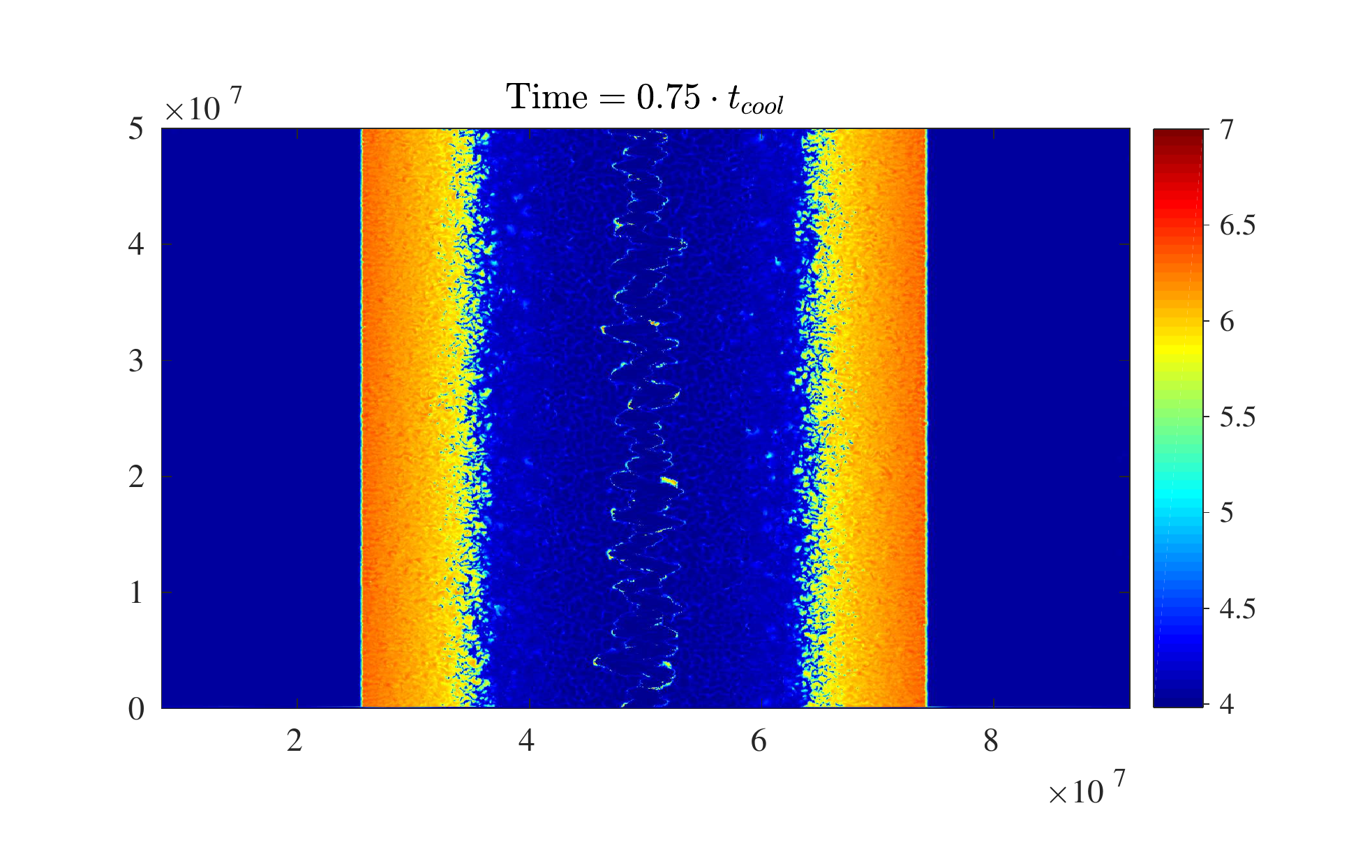}

\includegraphics[width=0.9\linewidth]{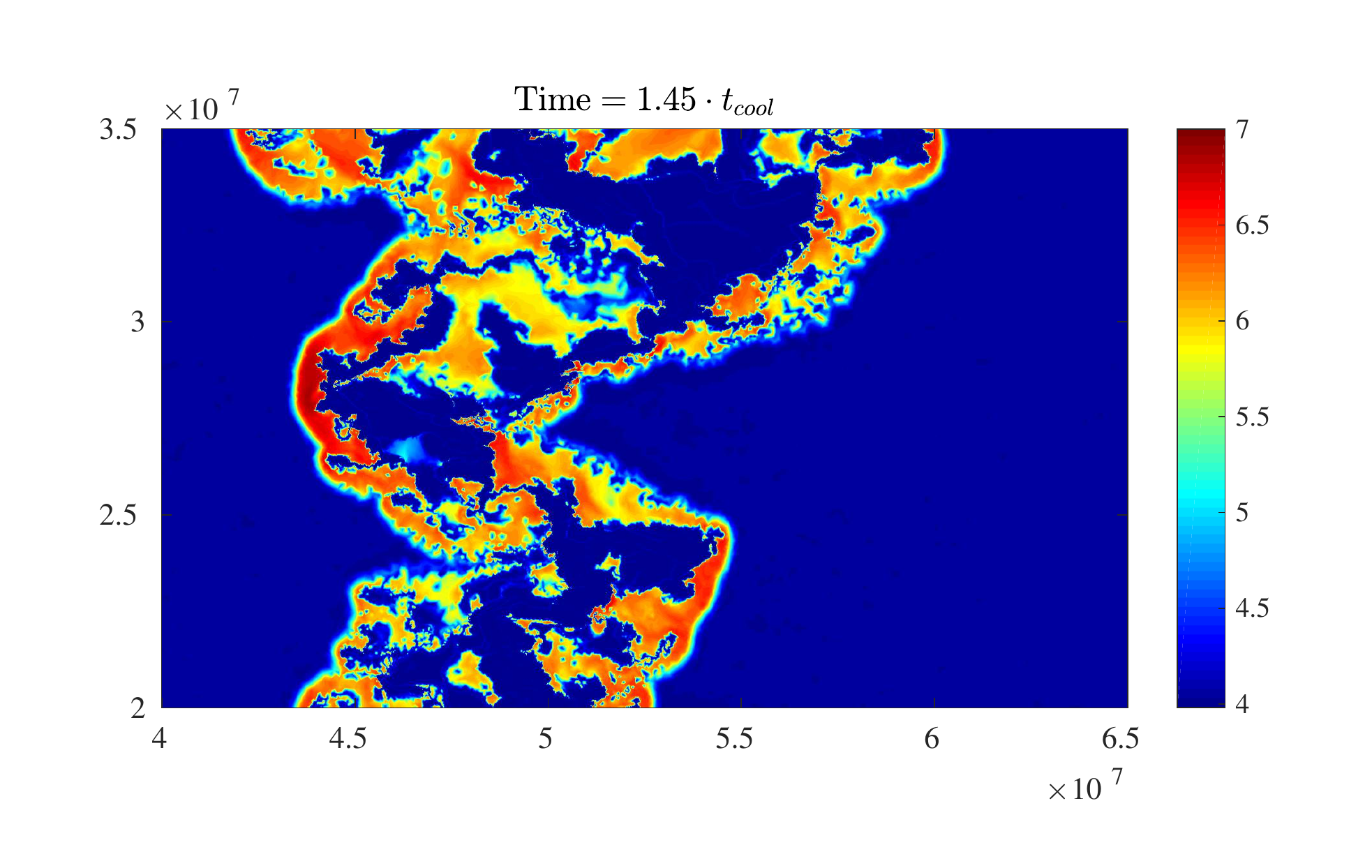}
\caption{Snapshots of the post-shock temperature, in units of log$_{10}$(T/K), for our fiducial model $M36\_L$ (``Lagrangian" scheme, Mach number $\mathcal{M} = 36$), shown times $t = 0.3t_{\rm cool}$ (top panel), $0.75t_{\rm cool}$ (middle panel), $1.45 t_{\rm cool}$ (bottom panel), where $t_{\rm cool}$ is the initial cooling timescale of the shocked gas.  The first snapshot shows the shock while it is still adiabatic and the shock front is stable, while the final snapshot shows the fully-developed non-linear thin shell instability.  Note the change in spatial scale of the final snapshot relative to the first two.}
\label{fig:walkthrough}
\end{figure}

The collision between opposite flows results in the creation of strong ``forward" and ``reverse" shocks (top panel of Fig. \ref{fig:walkthrough}), behind which the temperature is close to the value for a one-dimensional planar shock, $T_{\rm sh} \simeq 7\times 10^{6}$ K (eq.~\ref{eq:Tsh}).  After one cooling time has elapsed, the temperature at the center of the accumulating shell plummets from radiative losses (middle panel of Fig. \ref{fig:walkthrough}). This loss of pressure generates a rarefaction wave, which travels from the center out to the shock fronts. Once the rarefaction reaches the shocks, their loss of pressure support results in them being pushed back by the ram pressure of the upstream gas, collapsing around the cold dense central shell.  After the collapse, the upstream flows  now collide with the dense shell, which becomes unstable to the NTSI, driving the shock interface to an irregular shape.  

\begin{figure}
\centering
\includegraphics[width=0.9\linewidth]{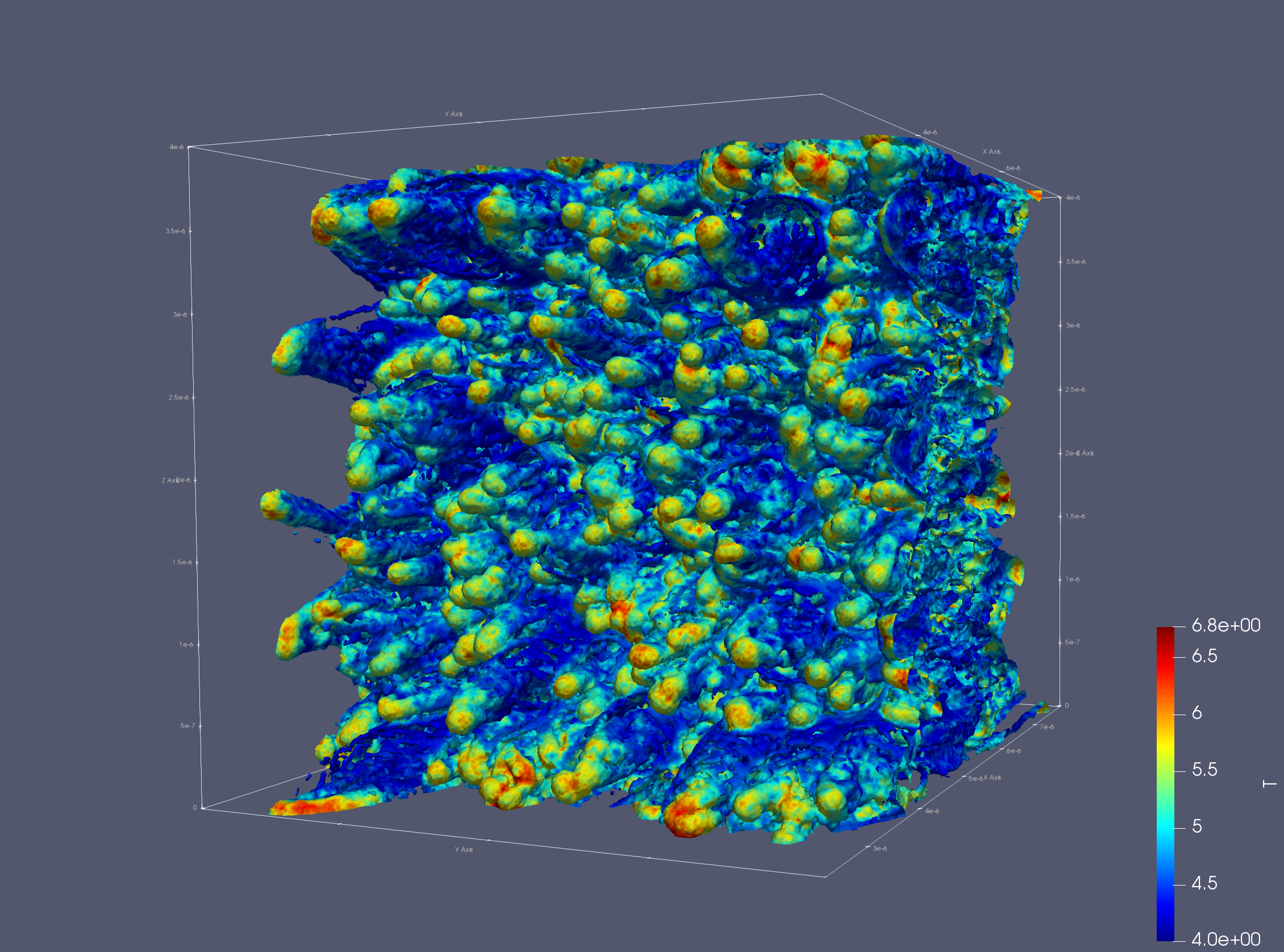}
\caption{Contour of the external shocked gas (set by density slightly larger than the inflow density) color coded by temperature for our 3D simulation. The NTSI corrugates the shock front and creates a "mountain range" topology.}
\label{fig:walkthrough3D}
\end{figure}

Due to the large bending angles in the central shell, most of the shocks are now oblique and, partially for this reason, produce gas with a temperature lower than $T_{\rm sh}$ (bottom panel of Fig. \ref{fig:walkthrough}). The NTSI quickly saturates, and the average properties of the shocked gas reach a steady state.  A movie of the evolution just described is available in the \href{https://www.eladsteinberg.com/movies}{online} version of the manuscript. In 3D the qualitative picture remains the same; Fig. \ref{fig:walkthrough3D} shows the temperature topology of the 3D shock front at saturation.

The above evolution qualitatively describes all of our simulations with sufficiently high Mach numbers (e.g.~$\mathcal{M} = 36$ for the example shown in Fig.~\ref{fig:walkthrough}).  For our lowest Mach number runs with $\mathcal{M} = 4$, the shock front largely remains laminar, a result explained by the fact that the NTSI  requires a minimum compression ratio to grow \citep{Vishniac94}.  Our lowest Mach number simulations also result in shocked gas which is thermally stable, which may explain some differences in the behavior.

\begin{figure}
\centering

\includegraphics[width=0.9\linewidth]{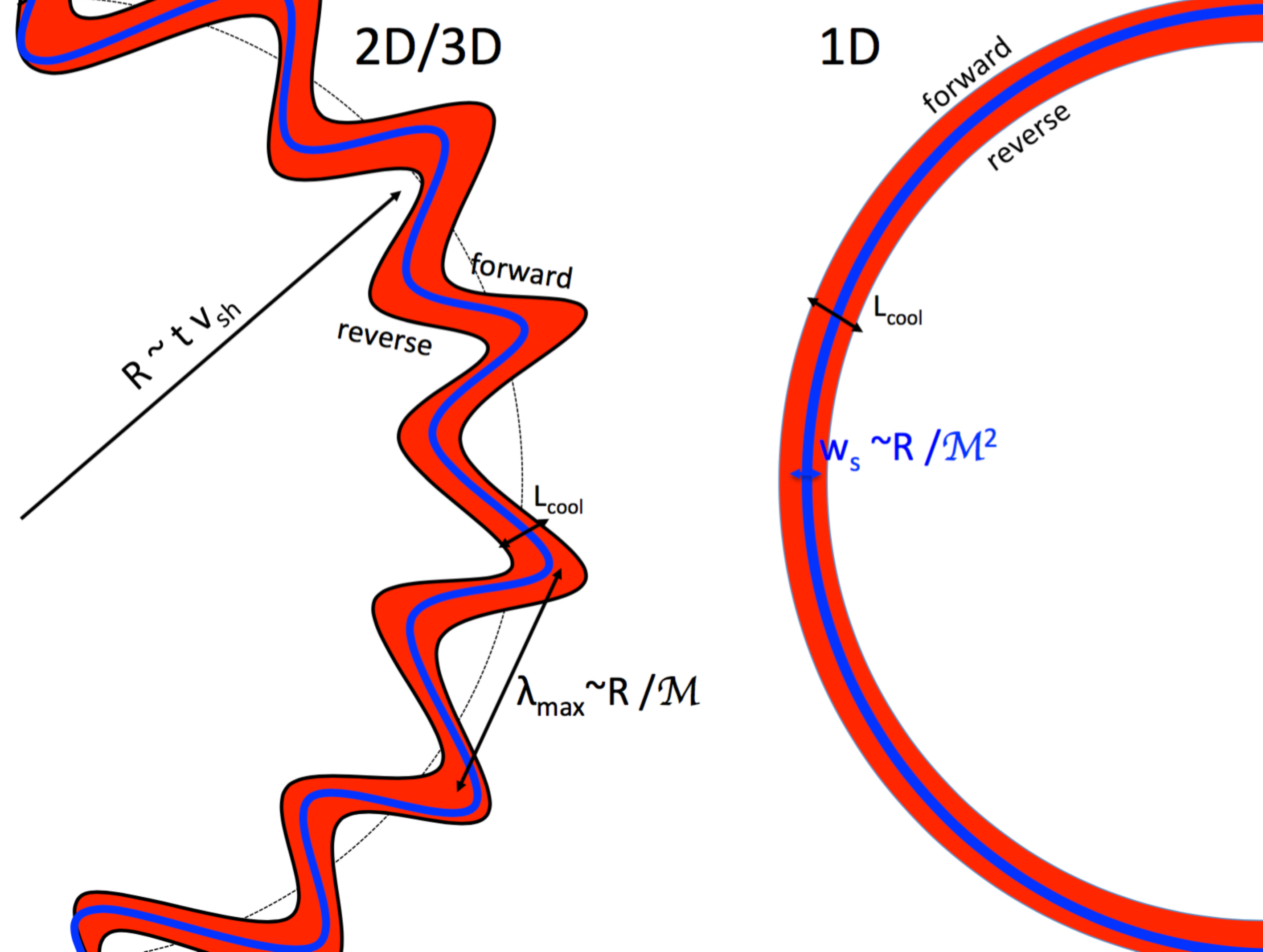}

\caption{Schematic illustration of the hierarchy of lengthscales $w_s \ll \lambda_{\rm max} \ll R$ in radially-expanding high-Mach number $\mathcal{M} \gg 1$ radiative shocks in an expanding medium with radial velocity comparable to the internal shock velocity $v_{\rm sh}$.  The radial scale of the flow on timescales $\sim t$ is $R \sim v_{\rm sh}t$.   In the naive 1D picture (shown on the right), the shell of cold gas bounded between the shocks is thin, with a radial width $w_s \approx c_{\rm s}t/\mathcal{M} \sim R/\mathcal{M}^{2}$.  However, in multiple dimensions (shown on the left), the shell is unstable to the NTSI, which in its saturated state creates structure in the shock front up to a characteristic lengthscale $\lambda_{\rm max} = c_{\rm s}t \sim R/\mathcal{M}$, where $c_{s}$ is the sound speed of the cold gas.  In both cases, the thickness of the layer of hot gas bounding the cold cold is approximately given by the cooling length $\mathcal{L}_{\rm cool}$ (eq.~\ref{eq:Lcool}).   }
\label{fig:hierarchy}
\end{figure}

Some aspects of the development of the NTSI in our simulations can also be understood from linear theory.  \cite{Vishniac94} shows that for an isothermal thin shell of width $w_s \approx c_{s}t/\mathcal{M}$, perturbations with wavelengths $\lambda$ in the range $w_s\le\lambda\ll c_s t$ are unstable\footnote{The maximum unstable wavelength $\lambda_{\rm max} = c_{\rm s}t$ results because the instability cannot grow supersonically, i.e.~$t^{-1}_{\rm NTSI}< c_s/\lambda$.}, and their amplitude, $A$, grows to saturation at a rate 
\begin{equation}
t^{-1}_{\rm NTSI}\approx \frac{A^{1/2}c_s}{\lambda^{3/2}}.
\end{equation}
In order for the perturbations to grow, an initial amplitude $ A \approx w_s$ is required.  At early times, the small wavelengths $\sim w_s$ grow the fastest, but as the width of the central shell increases, they become stable and are overtaken by the largest unstable wavelength $\lambda_{\rm max} = c_{s}t$.  The latter, which defines the largest length-scale characterizing the ``corrugated" shape of the shock front, is a factor of $\mathcal{M}$ times larger than the naive estimate of the shell thickness $w_s \approx c_{s}t/\mathcal{M}$ for a one-dimensional flow, but a factor $\sim 1/\mathcal{M}$ times smaller than the shock radius $R \sim v_{\rm sh}t$ for an expanding outflow of velocity $\sim v_{\rm sh}$.  This hierarchy of scales is schematically illustrated in Fig.~\ref{fig:hierarchy}.

\begin{figure}
\centering
\includegraphics[width=1.\linewidth]{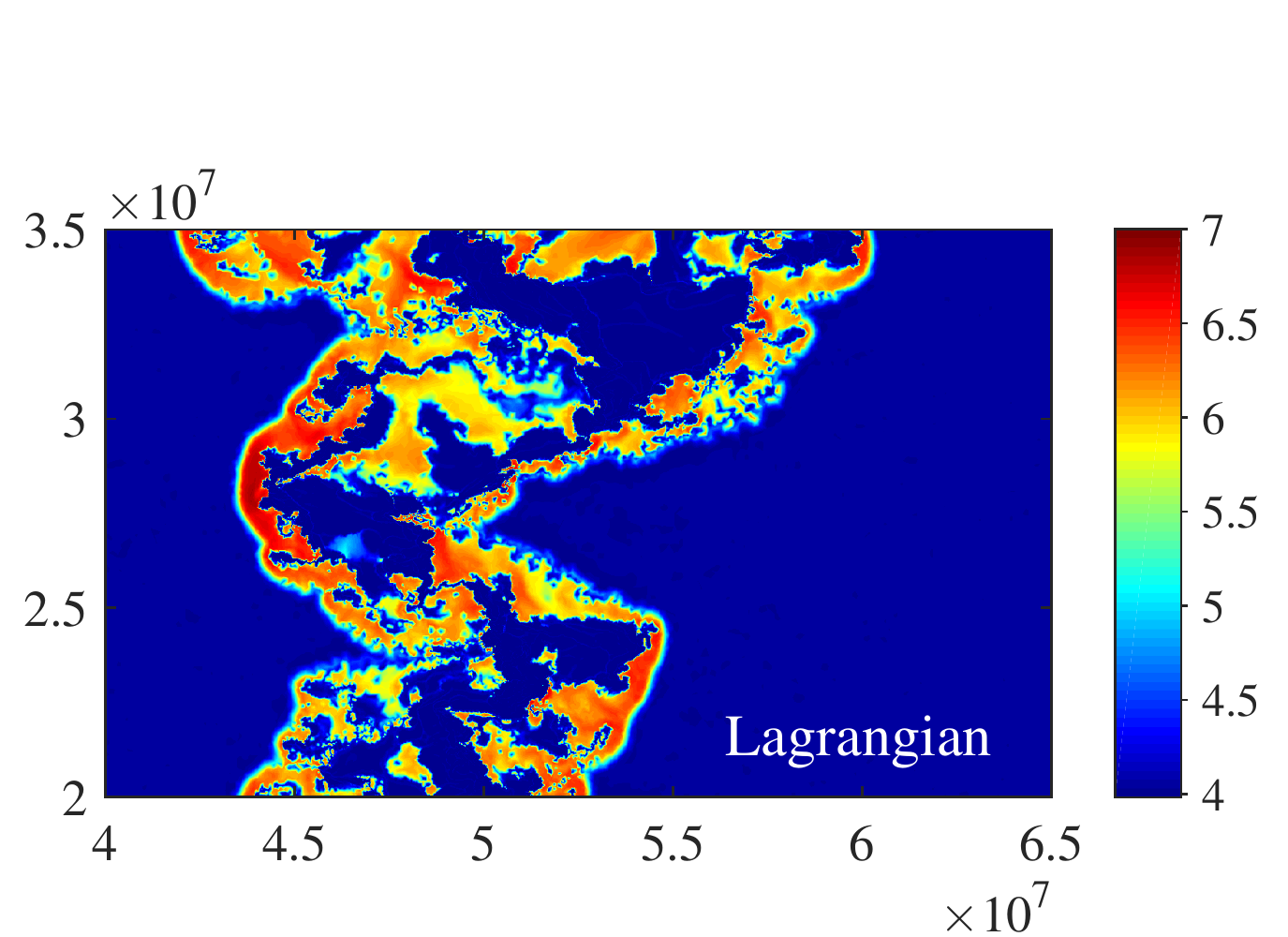}

\includegraphics[width=1.\linewidth]{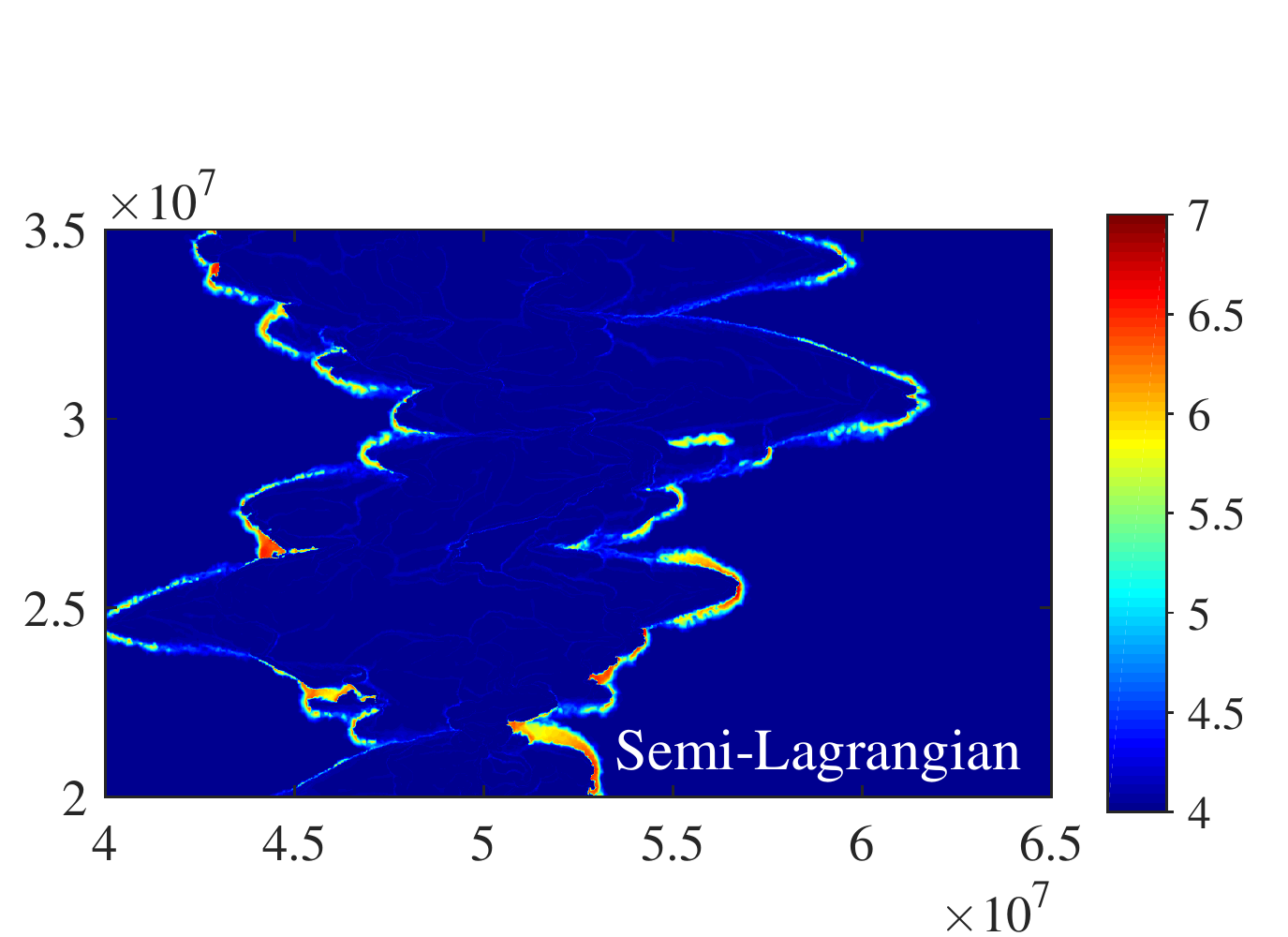}

\caption{Demonstration of how numerical conduction artificially suppresses the hot phase of the post-shock gas.  Here we show the temperature of the gas between the shocks, in units of log$_{10}$(T/K), in the saturated state of the NTSI.  We compare our fiducial model $M36\_L$ which adopts the ``Lagrangian" numerical scheme to minimize numerical conduction (top panel) to the ``Semi-Lagrangian" run $M36\_SL$ (bottom panel).}
\label{fig:tempcompare}
\end{figure}

\begin{figure}
\centering
\includegraphics[width=1.\linewidth]{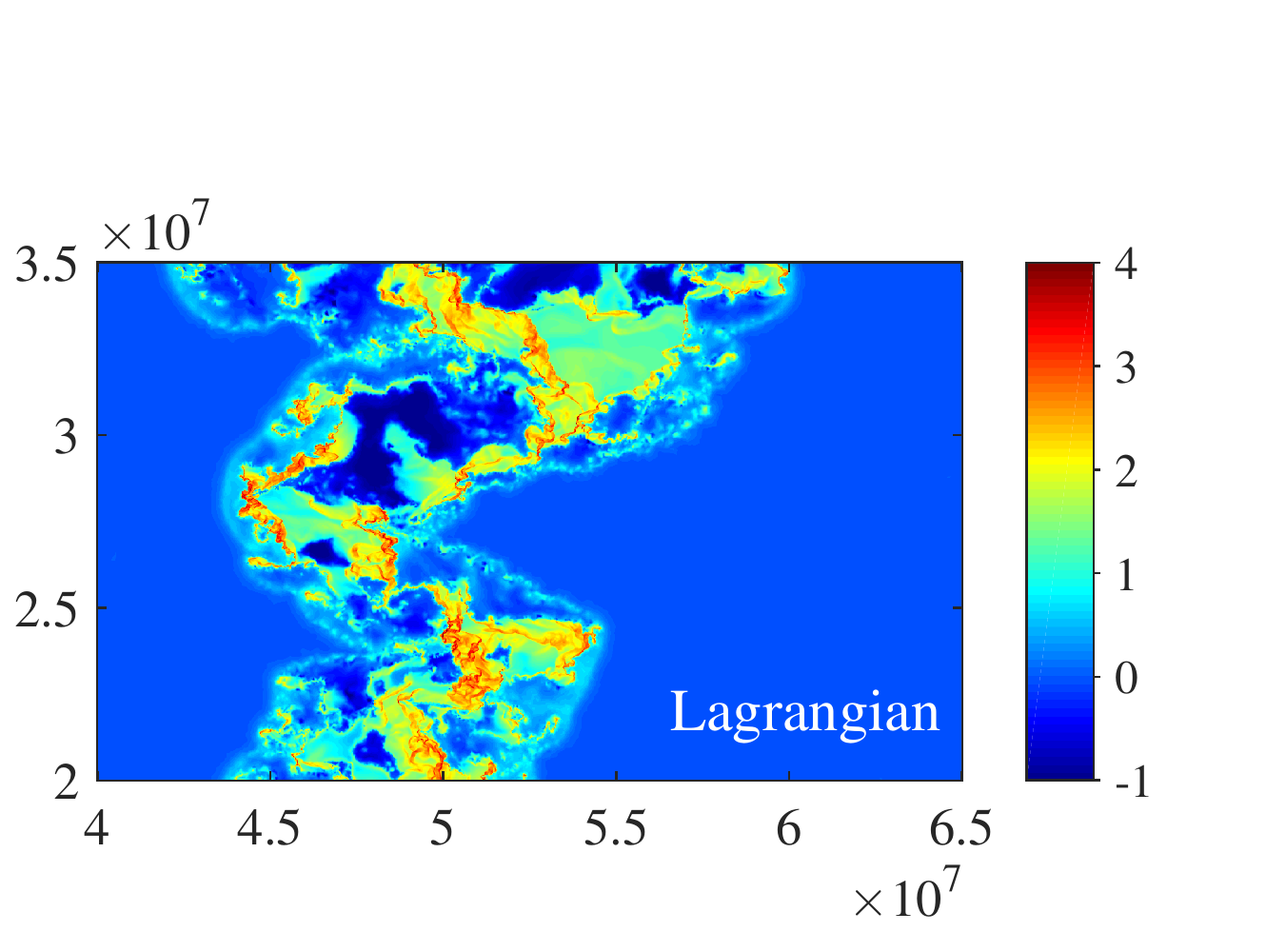}

\includegraphics[width=1.\linewidth]{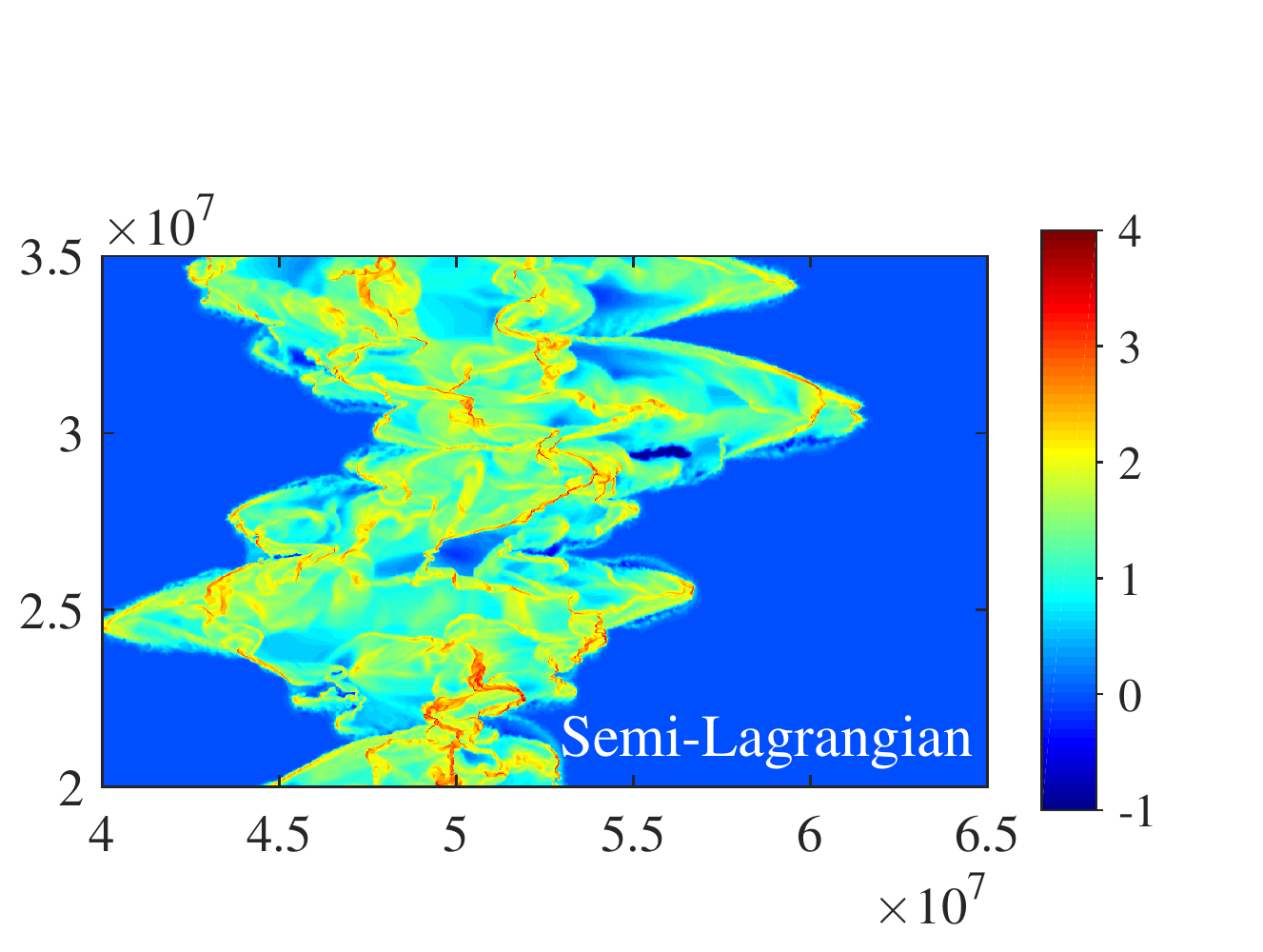}

\caption{Similar to Figure \ref{fig:tempcompare}, but comparing the density profiles of the $M36\_L$ and $M36\_SL$ models in the saturated state, in units of log$_{10}$ relative to the upstream density.}
\label{fig:densitycompare}
\end{figure}

For high Mach number shocks, radiative cooling results in an enormous density contrast $\rho_{\rm c}/\rho_{\rm h} \sim \mathcal{M}^{2}$ between the cold ($\sim 10^{4}$ K) and hot ($\sim 10^{6}-10^{7}$ K) phases of the post-shock gas (eq.~\ref{eq:contrast}), which can give rise to spurious cooling due to advection terms in the Euler equation (``numerical conduction").  To illustrate this effect, Figures \ref{fig:tempcompare} and \ref{fig:densitycompare} compare the temperature and density of the shocked gas after the NTSI has saturated between our ``Lagrangian" and ``semi-Lagrangian" simulations for the fiducial $\mathcal{M} = 36$ case.  In both schemes, rapid cooling and thermal instability in the post-shock gas creates filamentary cold structures.  However, in the ``Semi-Lagrangian" case artificial conduction of thermal energy from hot to cold cells (which radiate the received energy immediately because of the high cooling rate at low temperatures) effectively eliminates the hot phase between the shock fronts and thus artificially suppresses the temperature of the radiating gas ($\S\ref{sec:temperature}$).  In contrast, for our runs which employ the ``Lagrangian" technique, the post-shock region maintains separate cold and a hot phases, resulting in a higher overall temperature distribution.  Going to lower Mach numbers reduces the difference between the outcomes of the two schemes, to the limit where for $\mathcal{M} \simeq 4$ the two techniques give very similar results.

\subsection{Temperature Distribution of Post-Shock Gas}
\label{sec:temperature}

\begin{figure}
\centering
\includegraphics[width=0.9\linewidth]{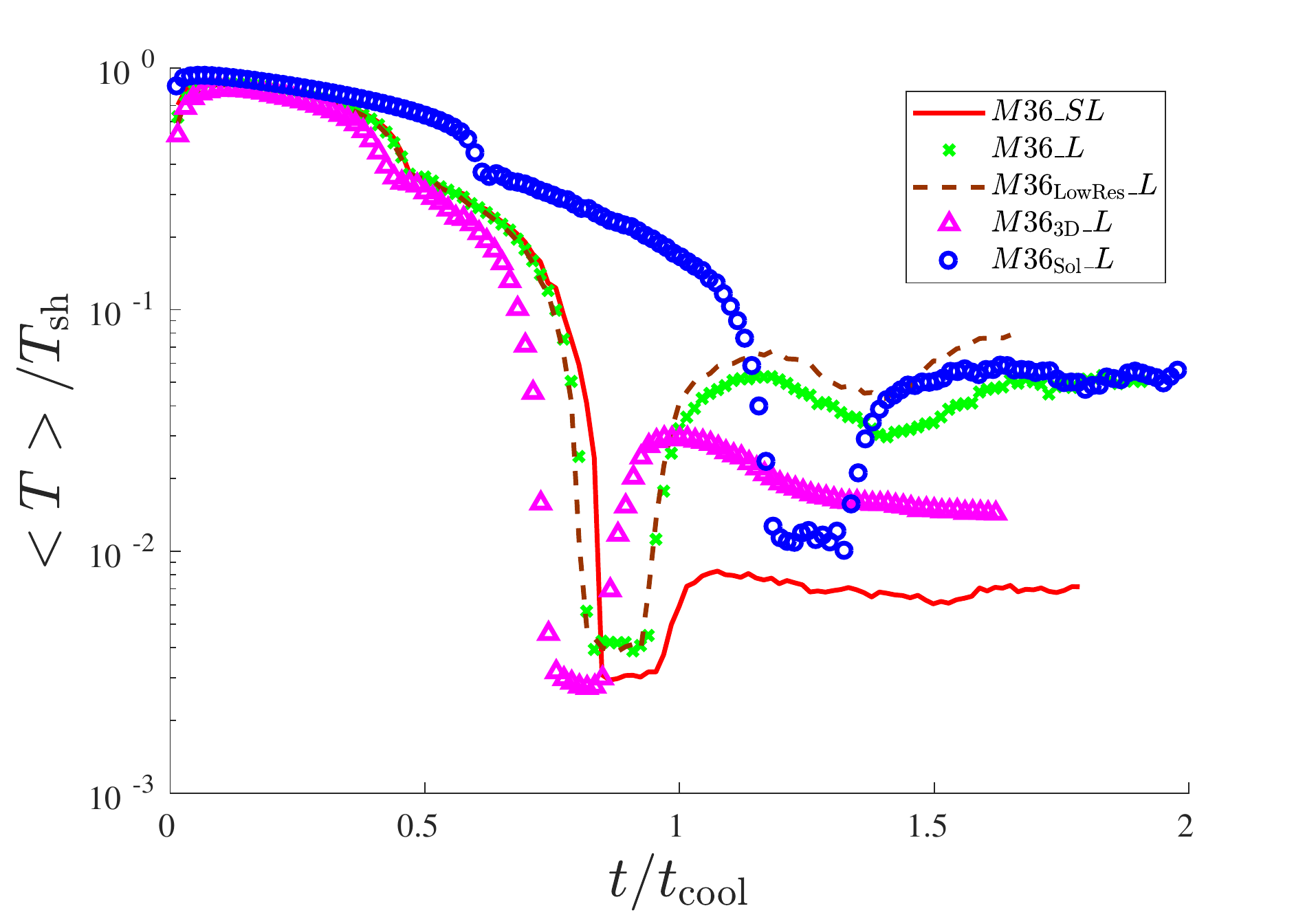}
\caption{Average temperature of the post-shock gas, weighted by the radiative luminosity and normalized to that for a head-one adiabatic shock $T_{\rm sh}$ (eq.~\ref{eq:Tsh}), as function of time in units of the initial cooling time, for the different simulation runs with Mach number $\mathcal{M} = 36$ (Table \ref{table:runs}).  This figure illustrates that the standard ``Semi-Lagrangian" simulation technique (red curve) produces much lower average post-shock temperatures than our new ``Lagrangian" technique (green curve; see also Fig.\ref{fig:tempcompare}), a result we attribute to the suppression of numerical conduction in the latter.  Also note that the 3D "Lagrangian" simulations (lavender curve) saturates to somewhat lower average temperatures than the otherwise equivalent 2D run.}
\label{fig:NTSI_T_M36}
\end{figure}

\begin{figure}
\centering
\includegraphics[width=0.9\linewidth]{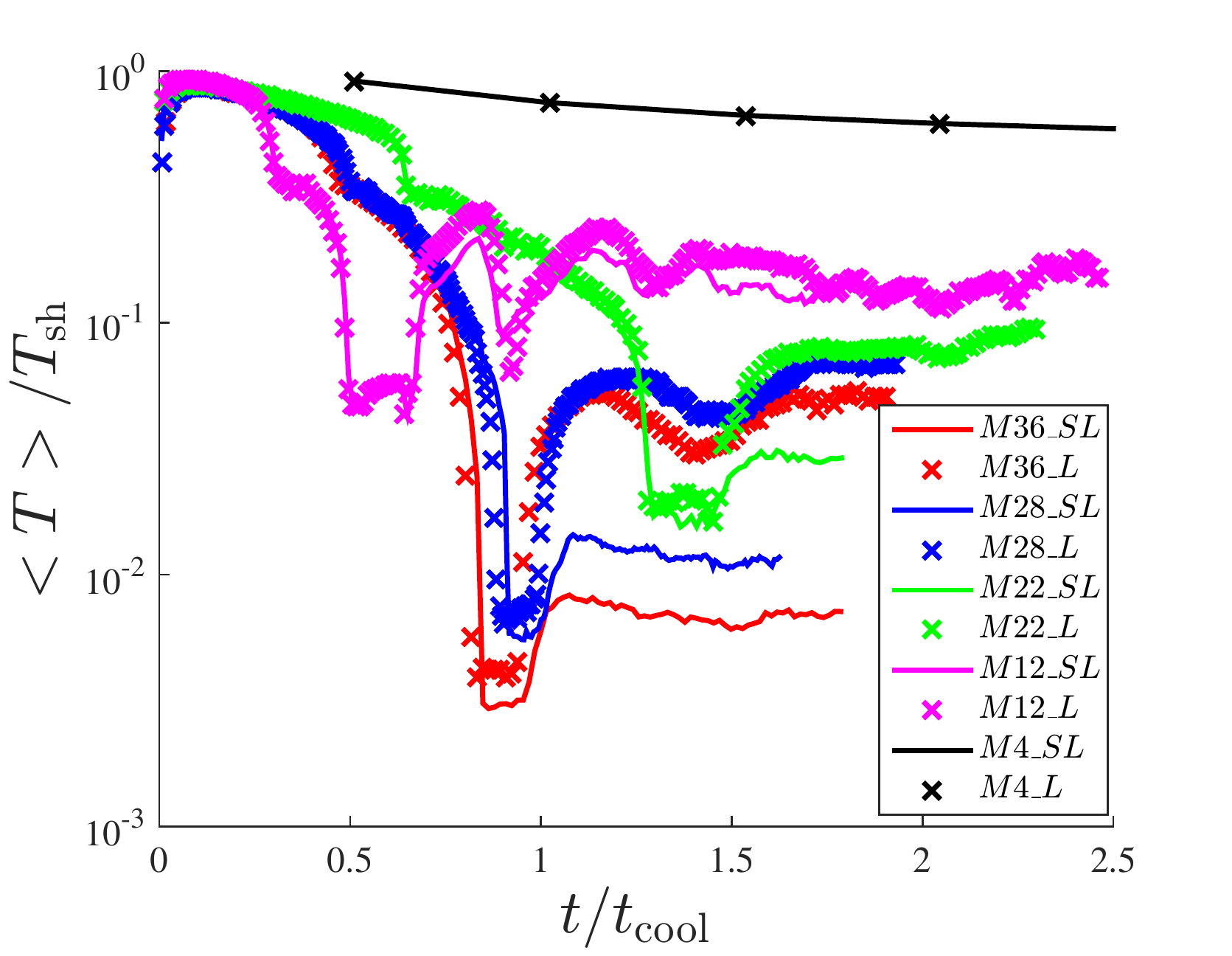}
\caption{Average temperature of the post shock gas, weighted by the radiative luminosity and normalized to that for a head-one adiabatic shock $T_{\rm sh}$ (eq.~\ref{eq:Tsh}), as function of time in units of the initial cooling time, for the different Mach number runs (Table \ref{table:runs}).}
\label{fig:NTSI_T_Mall}
\end{figure}

The geometrical structure of radiative shocks driven by thermal instabilities and the NTSI results in a reduction in the temperature of the radiating gas (e.g.~\citealt{Kee+14}).  Figures \ref{fig:NTSI_T_M36} and \ref{fig:NTSI_T_Mall} show for each of our simulations the time evolution of the average temperature, $\langle T\rangle$, weighted by the radiative luminosity.  We have normalized the latter to immediate post-shock temperature for a head-on adiabatic one-dimensional flow, $T_{\rm sh}$ (eq.~\ref{eq:Tsh}).  At early times ($t \ll t_{\rm cool}$), when the shock is still in the adiabatic state, we have $\langle T\rangle \approx T_{\rm sh}$, as expected.  For high Mach numbers however, the temperature plummets to a minimum after the shock collapses due to runaway cooling at times $t \sim t_{\rm cool}$.  Following this minimum, the temperature rises again slightly, eventually reaching an asymptotic value $\langle T\rangle$ which is typically one to two orders of magnitude smaller than $T_{\rm sh}$, depending on the Mach number.

Figure \ref{fig:NTSI_T_M36} shows the impact of numerical resolution and technique on the temperature in our results.  At early times, before cooling instabilities develop and while the post-shock structure is still well-resolved, our ``Semi-Lagrangian" simulations converge with resolution.  However, in the late-time saturated state of the NTSI, the average temperature of the ``Semi-Lagrangian" simulations is much lower than the otherwise equivalent ``Lagrangian" case (red versus green lines in Fig.~\ref{fig:NTSI_T_M36}).  The temperature furthermore increases with resolution, albeit slowly, and is therefore not converged for the ``Semi-Lagrangian" runs.  

By contrast, for the simulations which employ the ``Lagrangian" technique to prohibit advection between cells, the average temperature is almost an order of magnitude higher for high Mach numbers than even the highest resolution ``Semi-Lagrangian" simulations.  This implies that artificial (numerical) conduction plays a crucial role in the post-shock temperature distribution and confirms that our ``Semi-Lagrangian" simulations, and likely those employed in some past numerical works, are under-resolved.  Our ``Lagrangian" simulations also show evidence for convergence at late times, as shown by the similarity of the brown and green lines in Fig.~\ref{fig:NTSI_T_M36}.

\begin{figure}
\centering
\includegraphics[width=0.9\linewidth]{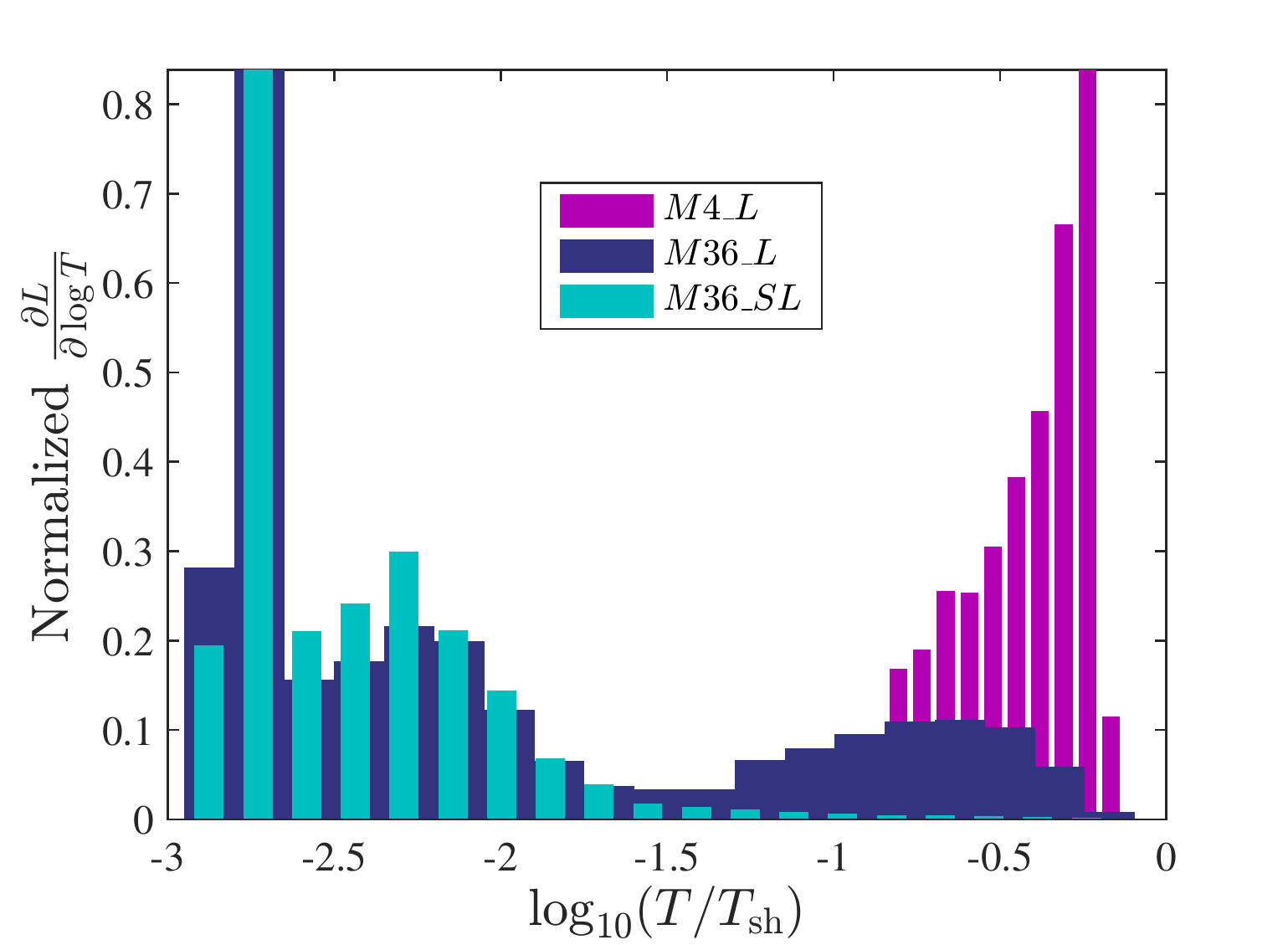}
\caption{Probability density function of the shock luminosity as a function of temperature, shown for a selected sample of runs from our suite of simulations (Table \ref{table:runs}). Low Mach number shocks are not susceptible to the NTSI and therefore dissipate most power at $T \approx T_{\rm sh}$, while for high Mach number the shock front becomes highly oblique and releases most of its energy at substantially lower temperatures.}
\label{fig:T_bar}
\end{figure}

Figure \ref{fig:T_bar} shows the probability density function of the radiated luminosity as function of the gas temperature in the saturated state for different runs. The ``Lagrangian" simulations again show a higher emission temperature than the otherwise equivalent ``Semi-Lagrangian'' case, whereas the lowest Mach run which is stable, emits all its luminosity at temperatures comparable to $T_{\rm sh}$. Table \ref{table:results} gives for each simulation the emission-weighted temperature and fraction of the radiated luminosity, $L(>T_{\rm sh}/3)/L_{\rm tot}$ emitted by cells at temperatures greater than $T_{\rm sh}/3$.  

\begin{figure}
\centering
\includegraphics[width=0.9\linewidth]{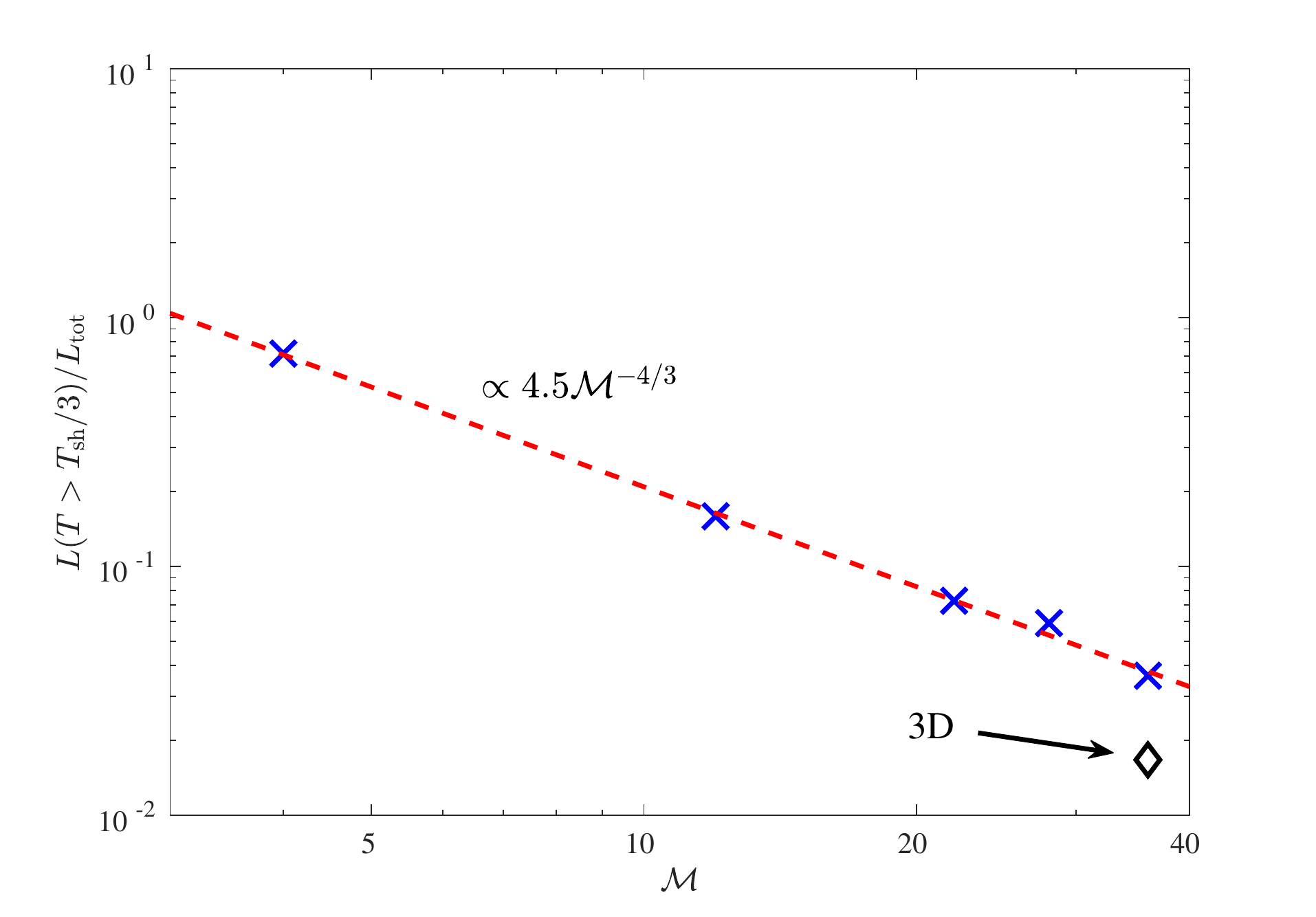}
\caption{Fraction of the total luminosity radiated behind the shock with X-ray temperature $T \gtrsim T_{\rm sh}/3$ as a function of the Mach number for our suite of Lagrangian simulations (Table \ref{table:runs}), where $T_{\rm sh}$ (eq.~\ref{eq:Tsh}) is the temperature for a one-dimensional adiabatic shock.}
\label{fig:T_reduction}
\end{figure}

Figure \ref{fig:T_reduction} shows the X-ray reduction as a function of the Mach number, which for our Lagrangian runs is reasonably fit by the functional form 
\be \frac{L(>T_{\rm sh}/3)}{L_{\rm tot}} \approx \frac{4.5}{\mathcal{M}^{4/3}}.
\label{eq:Kee}
\ee
This result is close to the conjecture of \citet{Kee+14} that the fraction of the radiation at X-ray emitting temperatures $\sim T_{\rm sh}$ is reduced by a factor that scales as $\sim 1/\mathcal{M}$ compared to the naive expectation of a one-dimensional flow.

\begin{figure}
\centering
\includegraphics[width=0.9\linewidth]{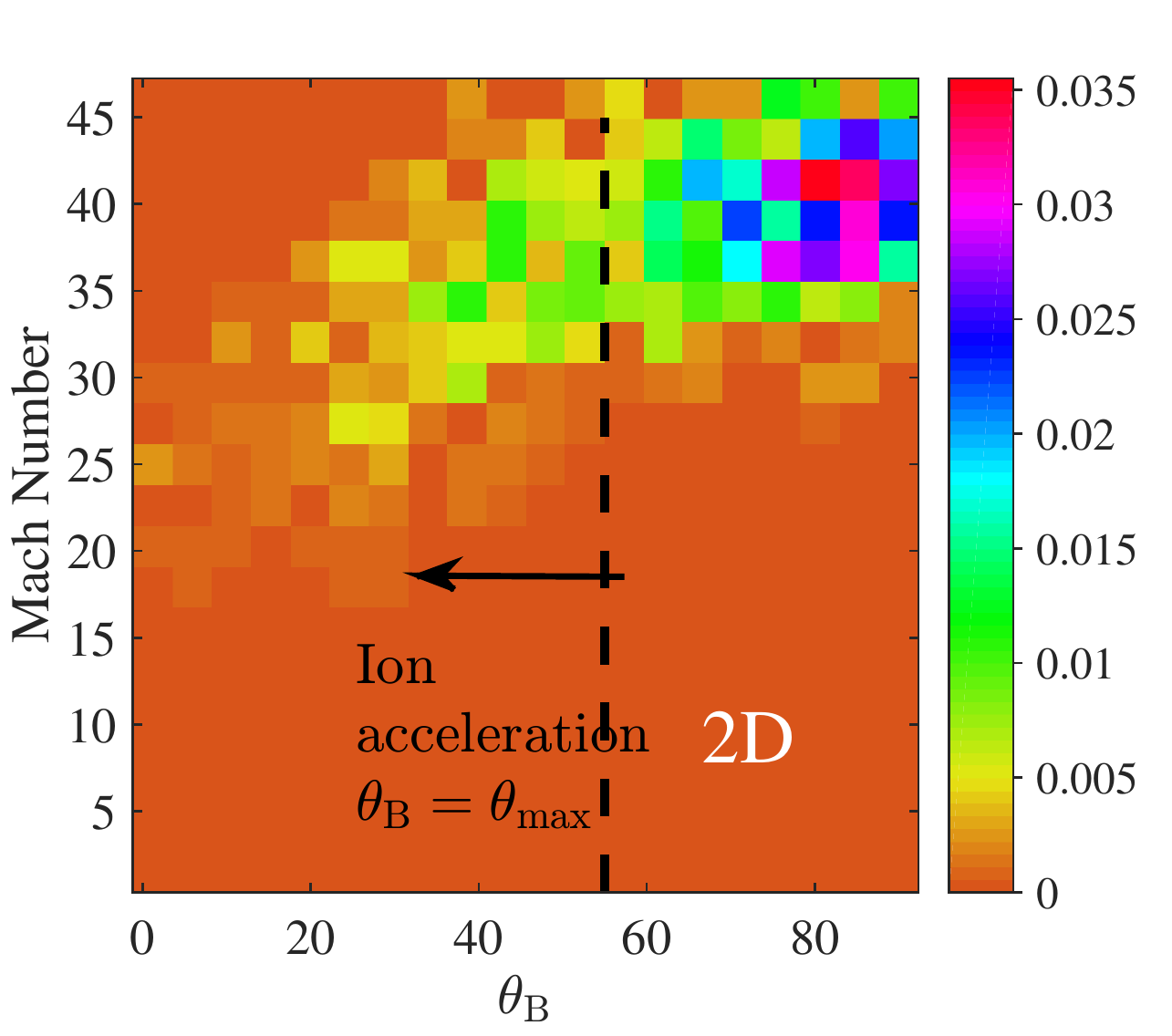}

\includegraphics[width=0.9\linewidth]{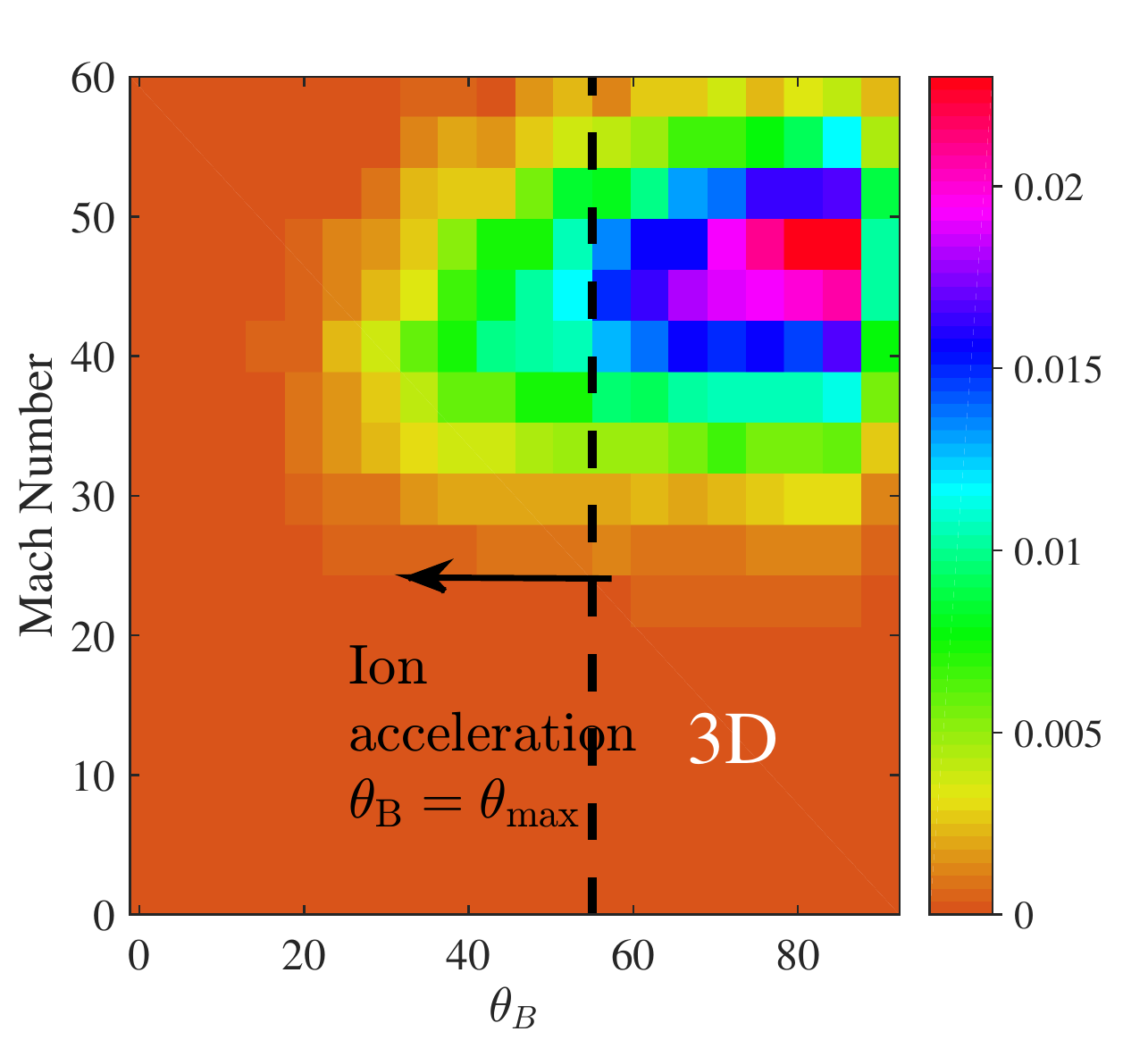}
\caption{Fraction of the upstream kinetic power which is dissipated across the radiative shock front as a function of the local Mach number and the angle $\theta_{\rm B}$ between the shock front and the assumed toroidal magnetic field, shown for our fiducial 2D run $M36\_L$ (top panel) and the equivalent 3D case $M36_{\rm 3D}\_L$ (bottom panel).  A vertical dashed line shows the critical angle, $\theta_{\rm max} \approx 40-55^{\circ}$, below which the shock accelerates relativistic ions with high efficiency \citep{Caprioli&Spitkovsky14}, under the assumption that the upstream magnetic field is perpendicular to the upstream velocity.}
\label{fig:Mach}
\end{figure}

Figure \ref{fig:Mach} shows a two-dimensional histogram of the shock power for our fiducial $\mathcal{M} = 36$ Lagrangian run as a function of Mach number and incident angle, $\theta_B$, between the shock front and the assumed tangential magnetic field (the Y axis). We find that our shock detection algorithm successfully detects $70-86\%$ of the incoming kinetic energy as being dissipated (converted into heat) at the external shock fronts, while the rest is dissipated in the turbulent region between the forward and reverse shock.  In the next section we describe how the inclination and Mach number distribution of the shocks plays an important role in the non-thermal ion acceleration.  
\begin{figure}
\centering
\includegraphics[width=0.9\linewidth]{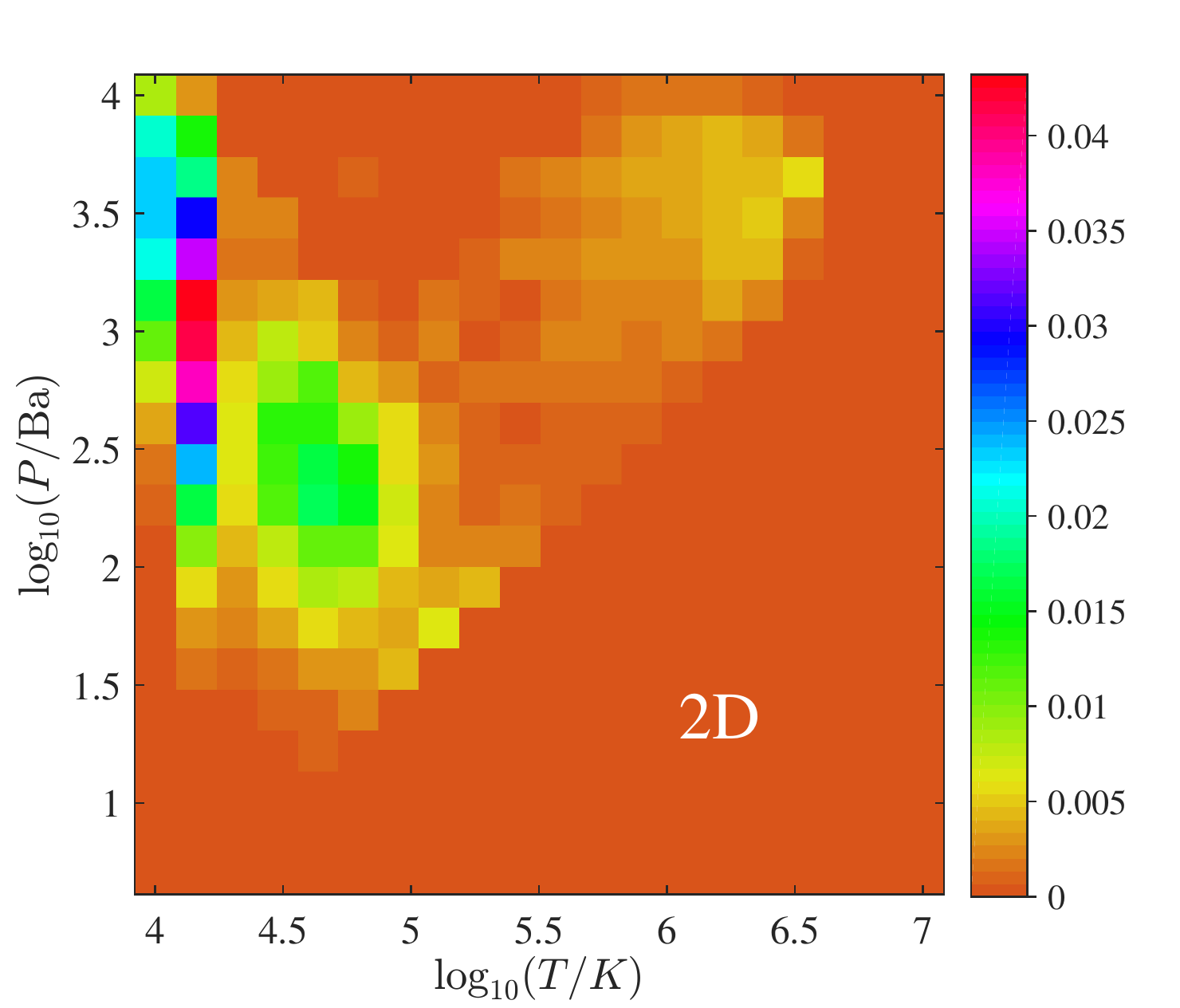}
\includegraphics[width=0.9\linewidth]{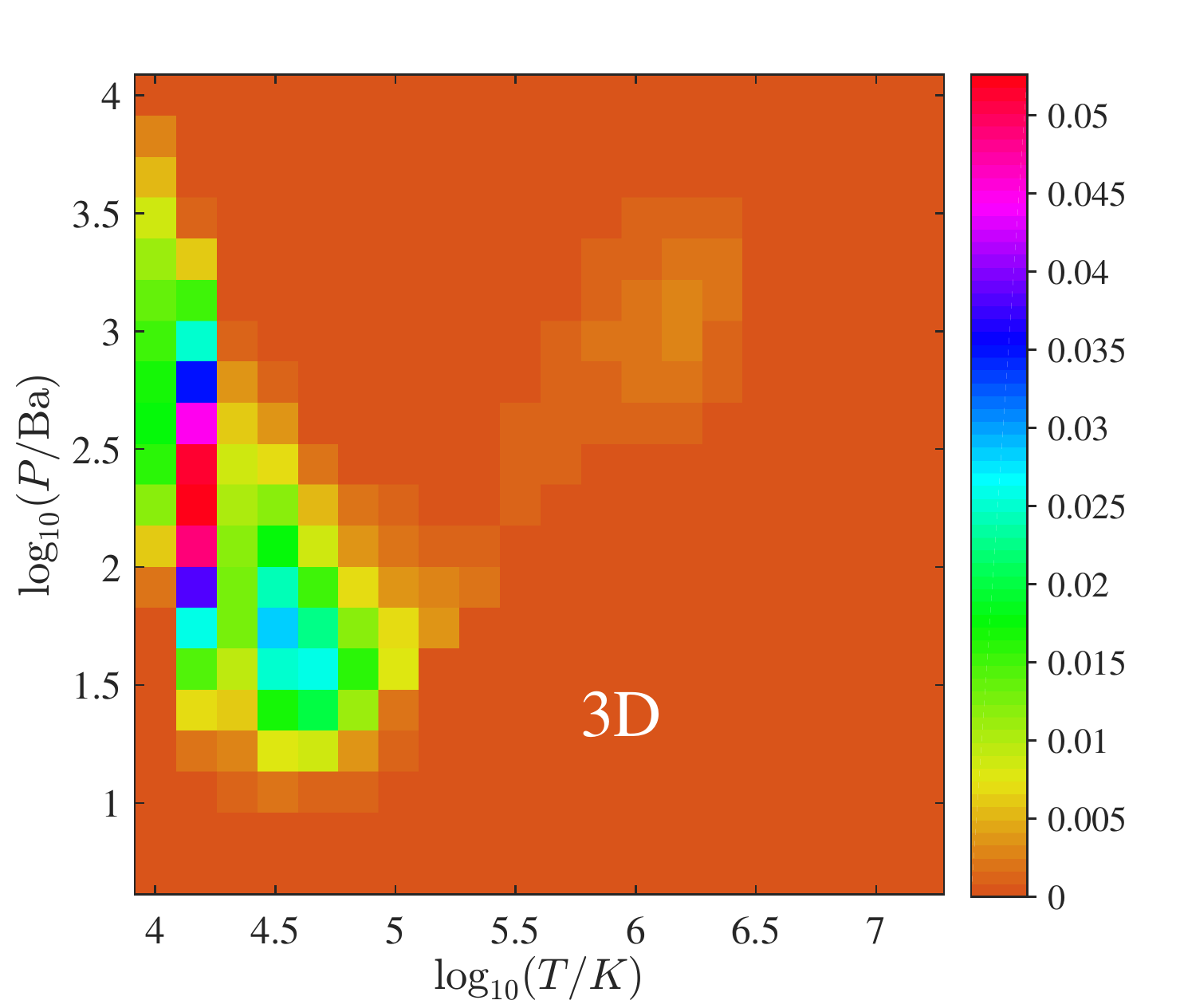}
\caption{Temperature and pressure distribution of the post-shock gas, weighted by its radiated luminosity, as measured at a snapshot in time following saturation of the NTSI.  The top panel shows the 2D run $M36\_L$, while the bottom panel shows the 3D case $M36_{\rm 3D}\_L$.  The hot gas has a higher pressure than much of the emitting cold gas, allowing the hot gas to transfer thermal energy to the cold gas via PdV work (e.g.~weak shocks) before it is radiated away.}
\label{fig:P-T}
\end{figure}

\begin{figure*}
\centering
\subfloat[Temperature(K)]{
\centering
\includegraphics[width=0.32\linewidth]{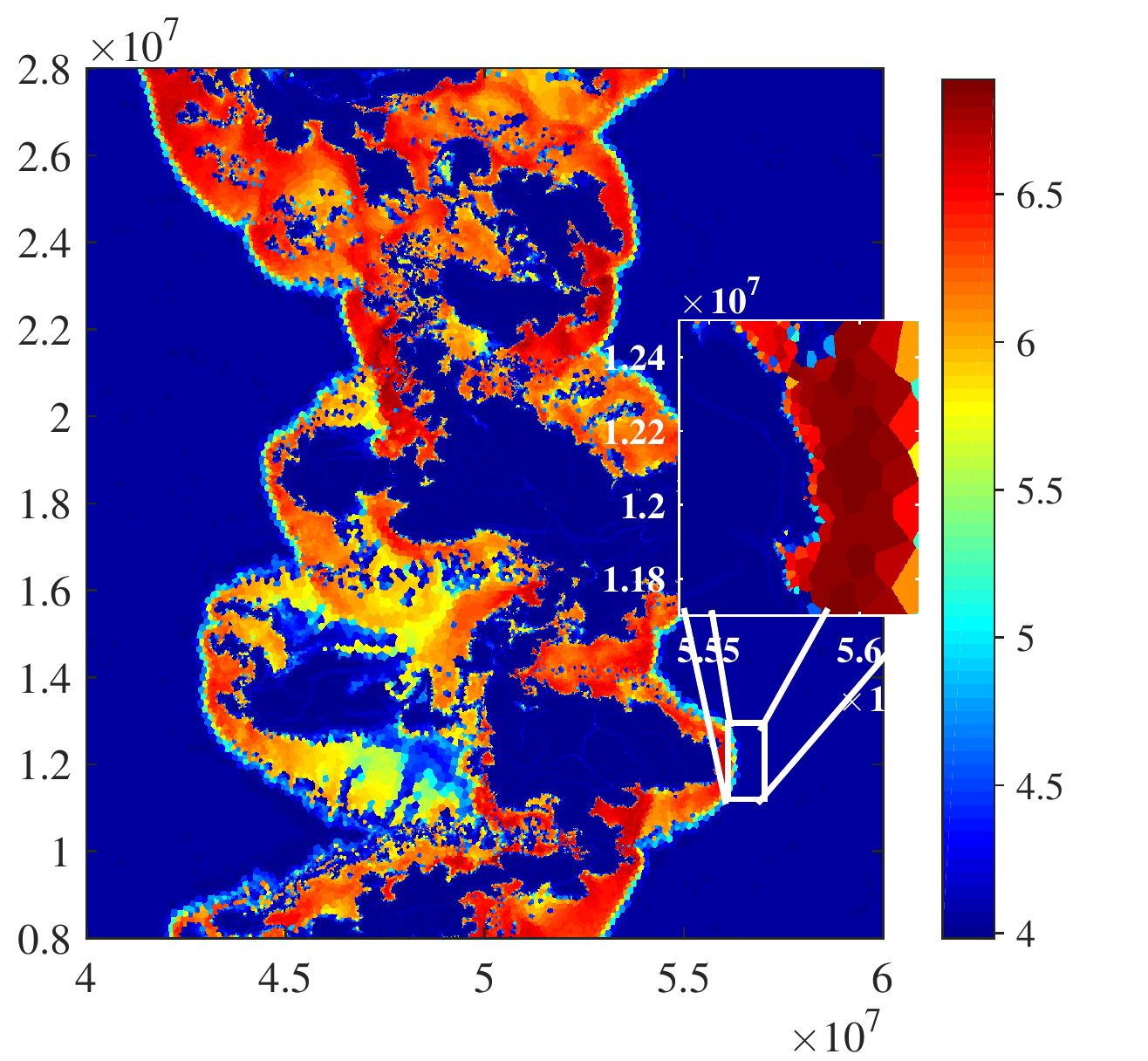}
}
\subfloat[Temperature(K)]{
\centering
\includegraphics[width=0.32\linewidth]{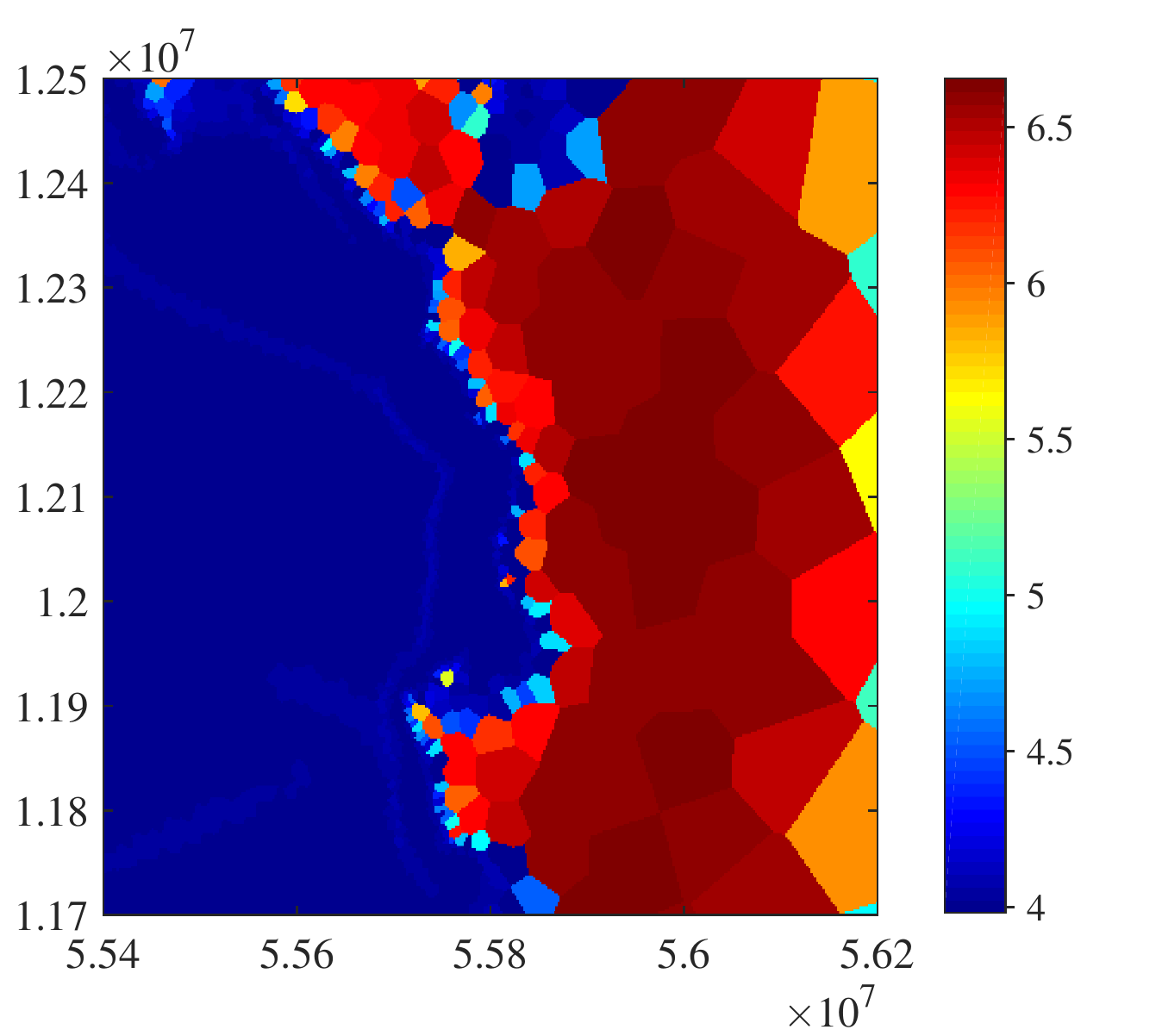}
}
\subfloat[Density(${\rm g/cm^3}$)]{
\centering
\includegraphics[width=0.32\linewidth]{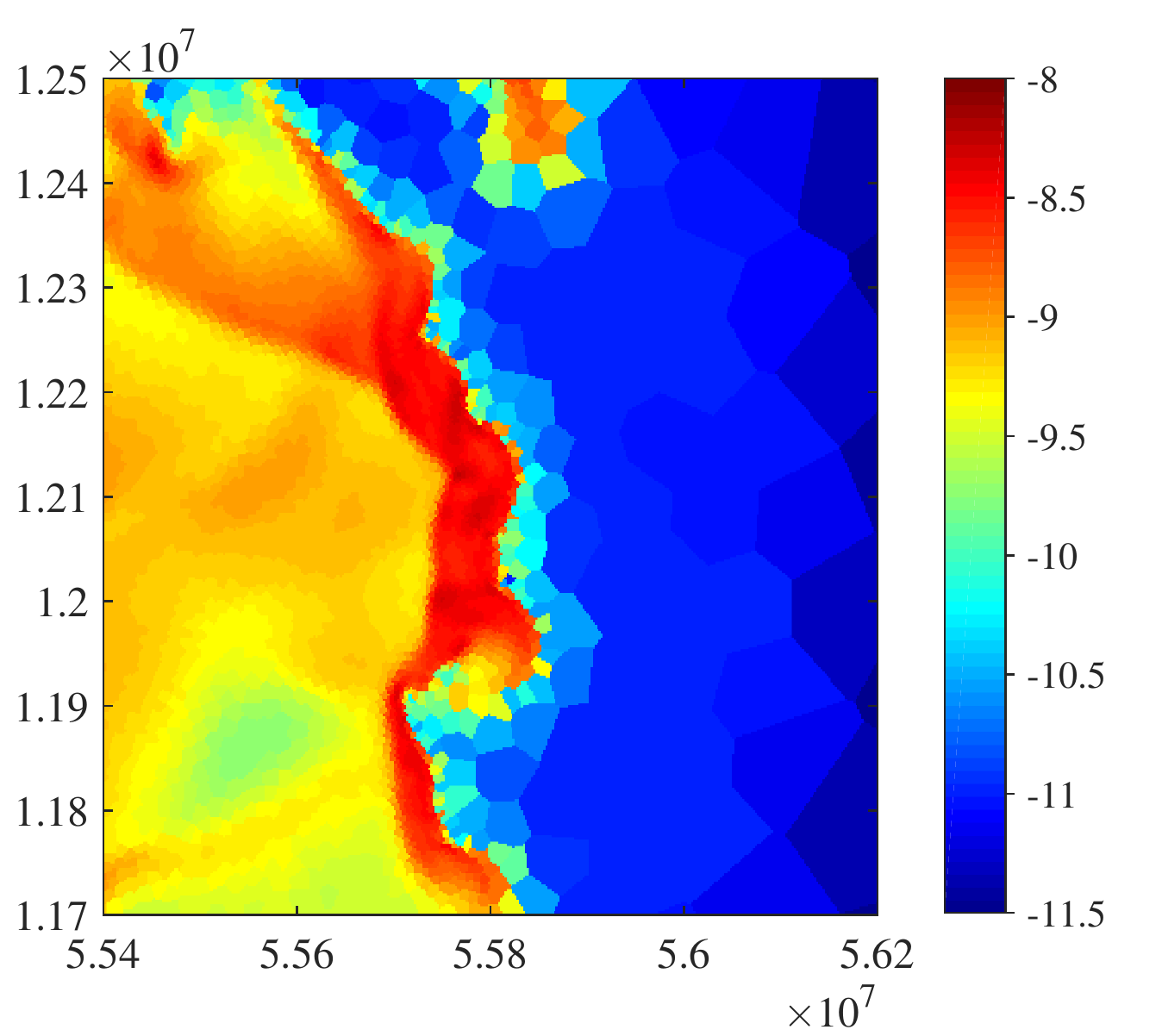}
}\\
\subfloat[Pressure(Ba)]{
\centering
\includegraphics[width=0.45\linewidth]{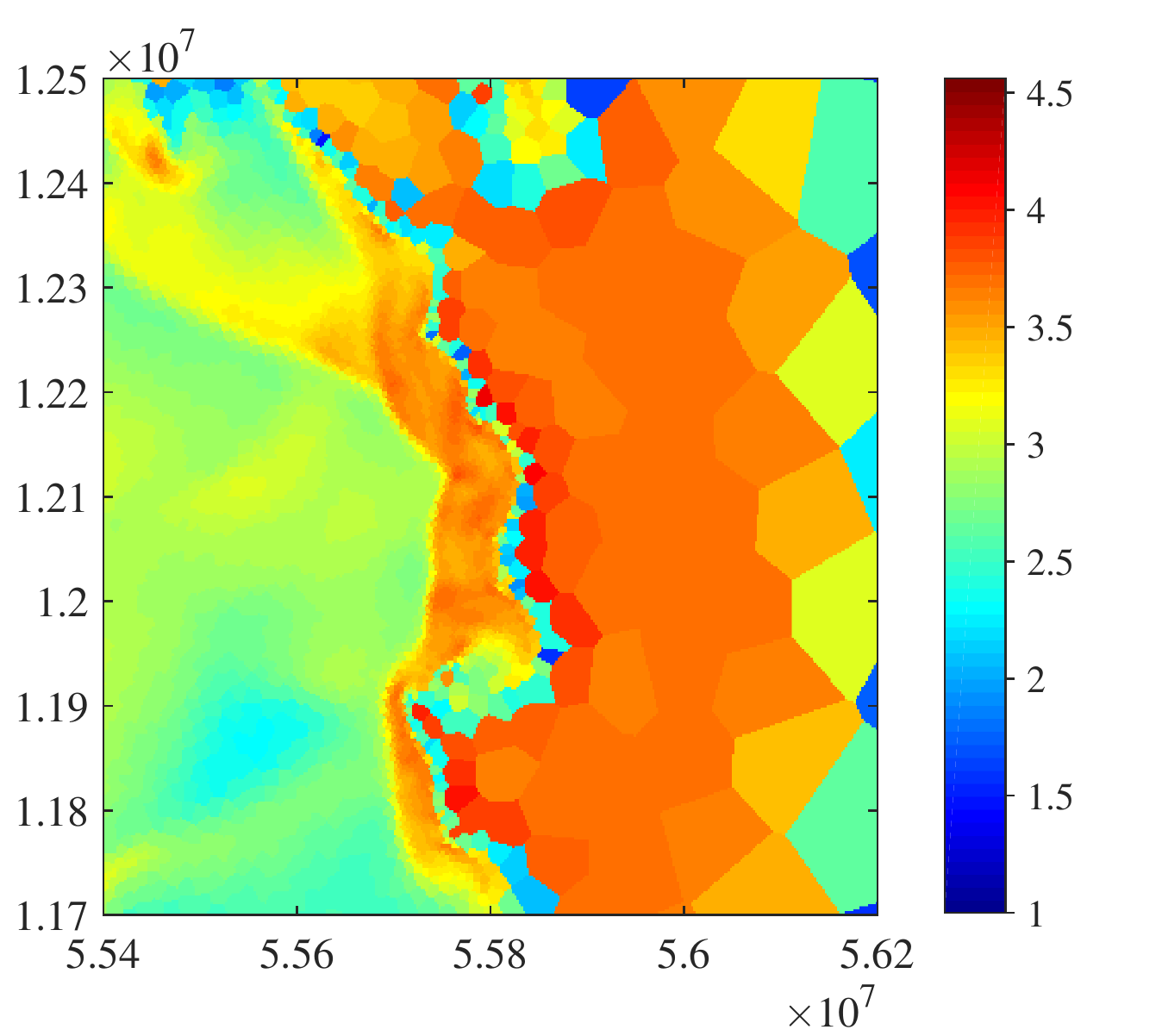}
}
\subfloat[Luminosity(erg/s)]{
\centering
\includegraphics[width=0.45\linewidth]{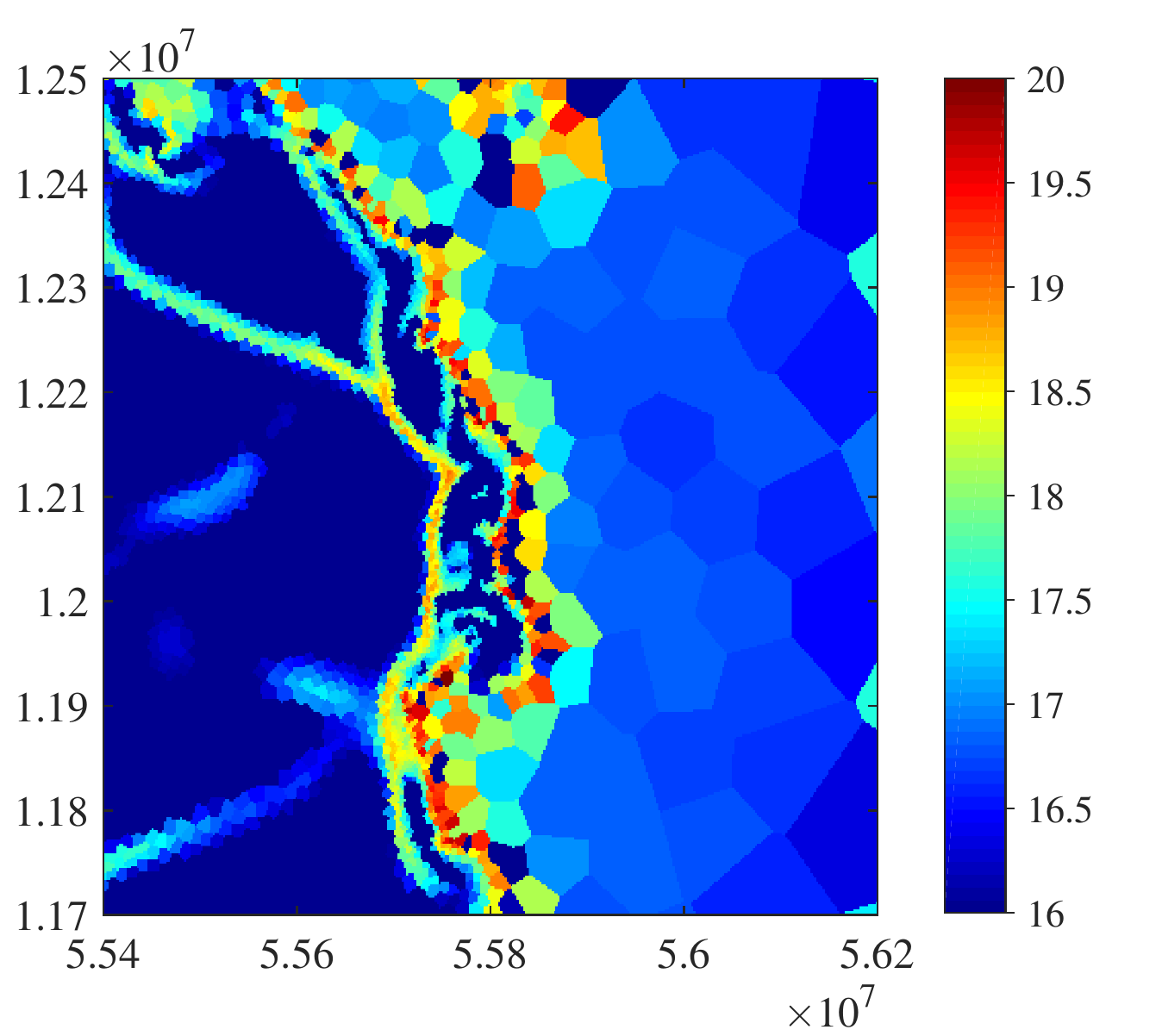}
}
\caption{Zoomed-in view of the interface between the outer (hot) shocked gas and the inner (cold) dense phase. The hot gas has a higher pressure and drives weak shocks into the cold dense phase, which is able to efficiently radiate the energy away before it can be radiated by the hot phase.  This effect is partially responsible for reducing the average temperature of the post-shock gas as compared to the naive expectation for a planar 1D compression analysis.}
\label{fig:zoom}
\end{figure*}

What physical effects cause the suppression in the X-ray luminosity of the shocks shown in Fig.~\ref{fig:T_reduction}?  Part of the reduction is related to the lower post-shock temperature resulting from the oblique shocks $T_{\rm sh} \propto \cos^{2}\alpha$; however, the distribution of shock power with angle shown in Fig.~\ref{fig:Mach} makes clear that this can hardly account for all of the suppression that we find (e.g., a factor of $\sim 30$ for $\mathcal{M} = 36$).  Some insight may be provided by Fig.~\ref{fig:P-T}, which shows the distribution of pressure and temperature of the post-shock gas, weighted by its radiated luminosity, for our fiducial 2D and 3D $\mathcal{M} = 36$ runs.  The faint patch at high $T$ and high $P$ represents the hot volume-filling post-shock gas, which cools radiatively towards the low $T$, low $P$ corner of the diagram.  Importantly, however, much of the cool emitting gas is under-pressurized relative to the hot surrounding medium.  This pressure difference enables the hot gas to perform P$\cdot$dV work on the cool gas, e.g. through a series of weak shocks, robbing the hot gas of its thermal energy, which is instead radiated with high efficiency at lower temperature.  To strengthen this conclusion, we show in Fig.~\ref{fig:zoom} a zoomed in region of the interface between the hot shocked gas and the cold dense gas. It is clear that the cells that are emitting most of the radiation are the cold, under-pressured cells adjacent to the hot shocked gas.  The under pressure we observe, which presumably results from runaway (acausal) cooling, appears to be a multi-dimensional effect, as it does not appear in one-dimensional simulations we have performed for the same shock parameters.

We now consider an order-of-magnitude analytic estimate of this effect.  The rate of shock heating of the cold gas by the hot gas is approximately given by
\be
L_{\rm c} \approx P_{\rm h}v_{\rm c}A_{\rm c} \sim  kT_{\rm h}n_{\rm h}A_{\rm c}v_{\rm sh}(\rho_{\rm h}/\rho_{\rm c})^{1/2},
\ee
where $P_{\rm h}$ is the pressure of the hot phase, $A_{\rm c}$ is the surface area of the cold gas, and $v_{\rm c}$ is the velocity of the shock driven into the cold gas by the hot phase.  The latter is estimated as $v_{\rm c} = v_{\rm sh}(\rho_{\rm h}/\rho_{\rm c})^{1/2}$ by equating the pressure of the hot gas $P_{\rm h}$ with the ram pressure $\rho_{\rm c}v_{\rm c}^{2}/2$ and using $P_{\rm h}/\rho_{\rm h} \approx v_{\rm sh}^{2}$.  The hot gas radiates the energy at a rate
\be
L_{\rm rad} \approx \frac{(3/2)kT_{\rm h}n_{\rm h}V_{\rm h}}{t_{\rm cool}} \sim kT_{\rm h}n_{\rm h}R^{2}v_{\rm sh},
\ee
where $V_{\rm h} = R^{2}\Delta$ is the volume of the hot gas across the shock front of area $R^{2}$, and $\Delta \approx \mathcal{L}_{\rm cool} = \eta R$ is the thickness set by the cooling length (eq.~\ref{eq:Lcool}).

Combining results, we conclude that
\be
\frac{L_{\rm c}}{L_{\rm rad}} \approx \frac{A_{\rm c}}{R^{2}}\left(\frac{\rho_{\rm h}}{\rho_{\rm c}}\right)^{1/2}.
\label{eq:Lcratio}
\ee
Thus, depending on to what extent the surface area of the (topologically complex) structure of cold filaments $A_{\rm c}$ exceeds the naive planar expectation $\sim R^{2}$, and on the ratio of the densities at the interface between the two media (Fig.~\ref{fig:zoom}), we can have $L_{\rm c} \gtrsim L_{\rm rad}$, i.e.~the shock work done by the hot phase on the cold gas can indeed be comparable to that directly radiated by the hot gas.  Since the cold filamentary gas has a larger surface area in 3D than in 2D, this could contribute to the lower average temperature of the radiation in our 3D simulations.

\subsection{Non-Thermal Ion Acceleration}
\label{sec:acceleration}

\begin{figure}
\centering
\includegraphics[width=0.9\linewidth]{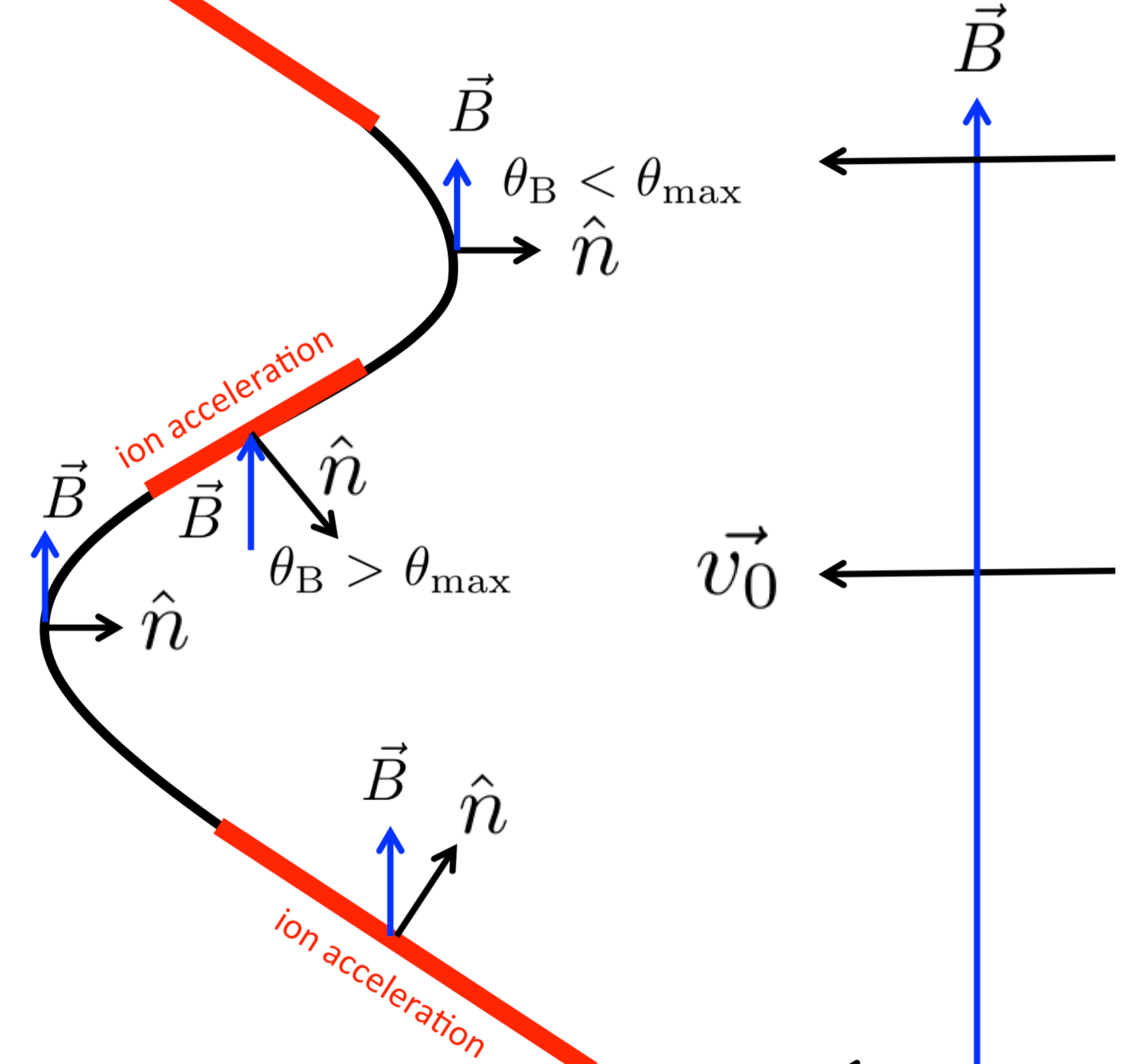}
\caption{Schematic diagram of the impact of the irregular geometry of the high Mach number radiative shock front on diffusive acceleration of relativistic ions.  Particle-in-cell simulations show that ion acceleration is limited to shocks for which the angle $\theta_{B}$ between the upstream magnetic field $\vec{B}$ and the shock normal $\hat{n}$ is less than a critical angle $\theta_{\rm max} \approx 40-55^{\circ}$ (``quasi-parallel" shocks; e.g.~\citealt{Caprioli&Spitkovsky14}).  In most astrophysical transients, due to magnetic flux conservation in the outflow, $\vec{B}$ is perpendicular to the radial upstream velocity $\vec{v}_{0}$ and therefore no acceleration can occur because for a planar shock $\hat{n}$ is everywhere parallel to $\vec{v}_{0}$.  However, the corrugated shock front which results from the NTSI creates localized regions where $\theta_{\rm B} < \theta_{\rm max}$ is satisfied (Fig.~\ref{fig:Mach}), resulting in a low but non-zero net ion acceleration efficiencies of $\epsilon_{\rm nt} \sim 0.01$ (Fig.~\ref{fig:e_nt}).}
\label{fig:schematicshock}
\end{figure}

Diffusive shock acceleration (DSA; e.g.~\citealt{Drury83,Blandford&Eichler87}) is the process by which a small fraction of charged particles crossing a shock front are accelerated to much higher energies than the thermal population by means of repeated reflections across the shock interface (usually due to their interaction with magnetic turbulence, often generated by the current of the non-thermal particles themselves).   

When active, linear DSA theory predicts that particles of mass $m$ and momentum $p$ are accelerated to a spectrum $f(p) \propto p^{-q}$, corresponding to an energy spectrum
\be
\frac{dN}{dE} = 4\pi p^{2}f(p)\frac{dp}{dE}.
\label{eq:dNdE}
\ee
In the non-relativistic regime $E = p^{2}/2m$, such that $dp/dE \propto E^{-1/2}$ and $dN/dE \propto E^{(1-q)/2}$; in the relativistic limit, $E \propto p$ and $dN/dE \propto E^{2-q}$.

The index $q$ is related to the compression factor $r$ according to
\be
q = \frac{3r}{r-1};\,\,\,\,\,r = \frac{(\gamma+1)\mathcal{M}^{2}}{(\gamma-1)\mathcal{M}^{2} + 2},
\label{eq:r}
\ee
where $\gamma$ is the adiabatic index.  In the limit of a strong shock $\mathcal{M} \gg 1$, we have $r = q = 4$ and the spectrum at high energies is a power-law $dN/dE \propto E^{-\beta}$ with $\beta = q-2 \approx 2$.  However, for lower $\mathcal{M}$ as obtained in the oblique shocks induced by the NTSI, $r \lesssim 4 (q \gtrsim 4)$, and we expect $\beta \gtrsim 2$.  Values of $q > 4$ can also result from non-linear effects at the shock due to the feedback of cosmic rays on the compression factor (e.g.~\citealt{Malkov&Drury01}), but we ignore this effect here since the acceleration efficiencies we find are quite modest.

The power-law index $\beta$ of the non-thermal particle energy distribution depends on the Mach number of the shock.  However, the overall efficiency $\epsilon_{\rm nt}$ with which the kinetic power of the shock is converted into non-thermal particle energy instead depends mainly on the orientation of the magnetic field ahead of the shock.  \cite{Caprioli&Spitkovsky14} show that $\epsilon_{\rm nt}\approx 0.05-0.2$ for upstream magnetic fields quasi-parallel to the shock normal (depending weakly on the Mach number), but that $\epsilon_{\rm nt}$ rapidly decreases for magnetic field inclinations $\theta_{\rm B} \gtrsim 40^{\circ}$, approaching $\epsilon_{\rm nt}\approx 0$ for $\theta_{\rm B} \gtrsim \theta_{\rm max} = 40-55^{\circ}$.  

Because in the transient astrophysical systems of interest the upstream magnetic field is expected to be nearly perpendicular to the radially-propagating shock fronts (i.e. $\alpha - 90^{\circ} = \theta_{\rm B}$), little or no particle acceleration is expected if the shock front remains smooth, in conflict with observations of GeV gamma-rays from classical novae.  However, the corrugated shock structure created by the NTSI for high Mach number results in localized regions where $\theta_{\rm B} \lesssim \theta_{\rm max}$ (Fig.~\ref{fig:Mach}), as illustrated schematically in Fig.~\ref{fig:schematicshock}, resulting in a smaller but non-zero average particle acceleration efficiency.  The impact of fluctuations in the direction of the upstream magnetic field relative to the shock normal on the injection of ions into the DSA cycle has been explored in previous works (e.g.~\citealt{Guo&Giacalone10}); however, the instabilities explored here provide a natural and self-consistent mechanism for creating the requisite geometry.        

\begin{figure}
\centering
\includegraphics[width=0.9\linewidth]{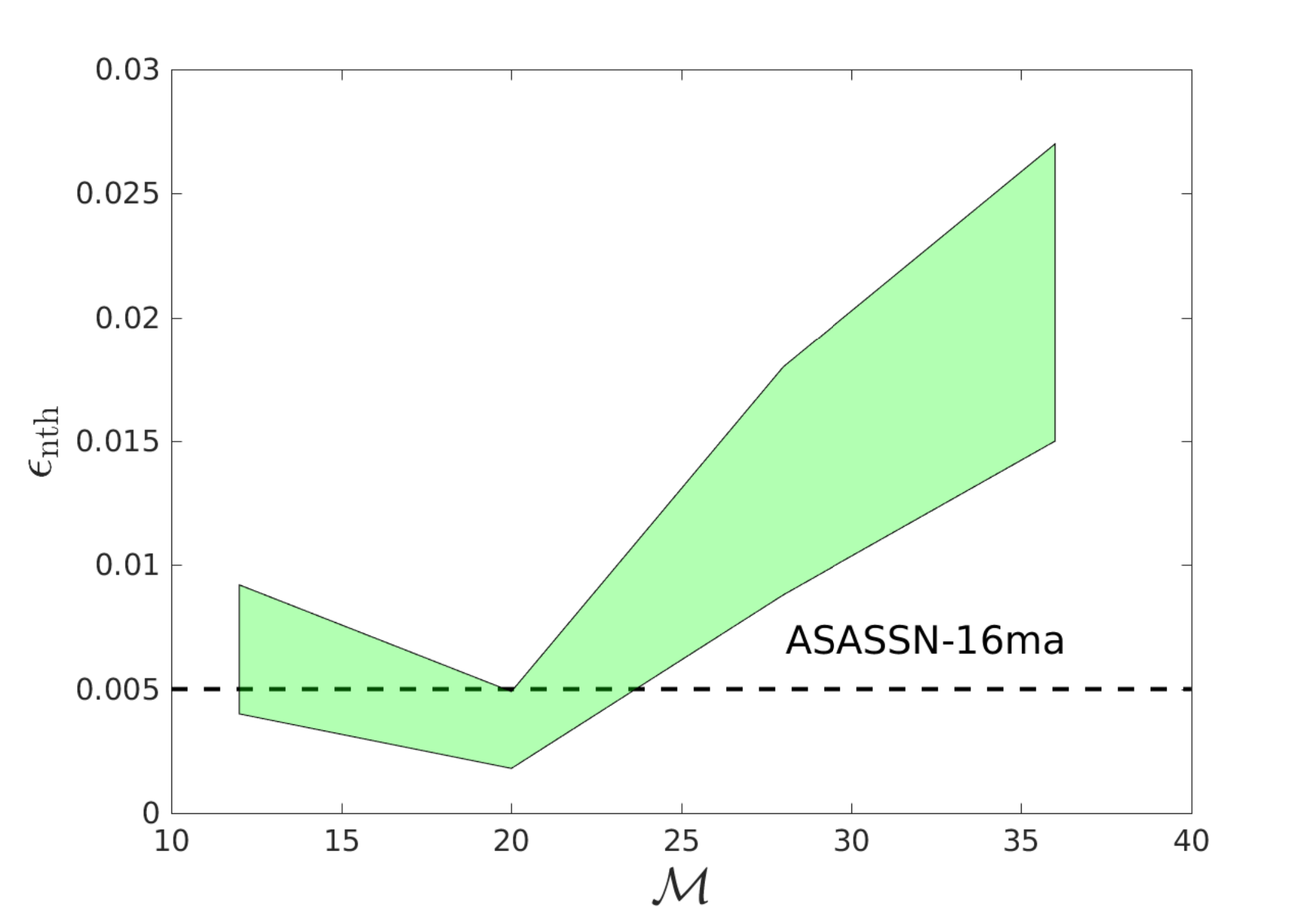}
\caption{Average non-thermal particle acceleration efficiency, $\epsilon_{\rm nt}$, of radiative shocks as a function of the Mach number. The upper border of the shaded region is given by a maximum angle of $55^\circ$ between the shock normal and the upstream magnetic field and the lower border is given by $40^\circ$. Also shown for comparison is the derived efficiency from calorimetry applied to the gamma-ray emitting classical nova ASASSN-16ma \citep{Li+17}.}
\label{fig:e_nt}
\end{figure}

Figure \ref{fig:e_nt} shows our calculation of the non-thermal particle acceleration efficiency, $\epsilon_{\rm nt}$, as a function of Mach number.  These are obtained by time-averaging the output for each of the numerical simulations and weighing the kinetic power of the shocked cells which satisfy the $\theta_{\rm B} \lesssim \theta_{\rm max}$ condition of \citet{Caprioli&Spitkovsky14}.  For concreteness, we approximate the results of the latter by assuming $\epsilon_{\rm nt} = 0.1$ for $\theta_{\rm B} \le 55^{\circ}$ and $\epsilon_{\rm nt} = 0$ for $\theta_{\rm B} > 55^{\circ}$; however, our results are not  sensitive to the precise transition angle (varying by a factor $\lesssim 2$ if we instead take $\theta_{\rm max} = 40^{\circ}$; Table \ref{table:results}).  Additionally, we assume that the magnetic field is not altered during the passage of the shock wave. We find that the acceleration efficiency $\epsilon_{\rm nt}$ is roughly $\approx 0.01$, depending only weakly on Mach number provided the latter is large enough for growth of the NTSI.  This value agrees remarkably well with the ion acceleration efficiency inferred from modeling the optical and gamma-ray emission from ASASSN-16ma \citep{Li+17}.

\begin{figure}
\centering
\includegraphics[width=0.9\linewidth]{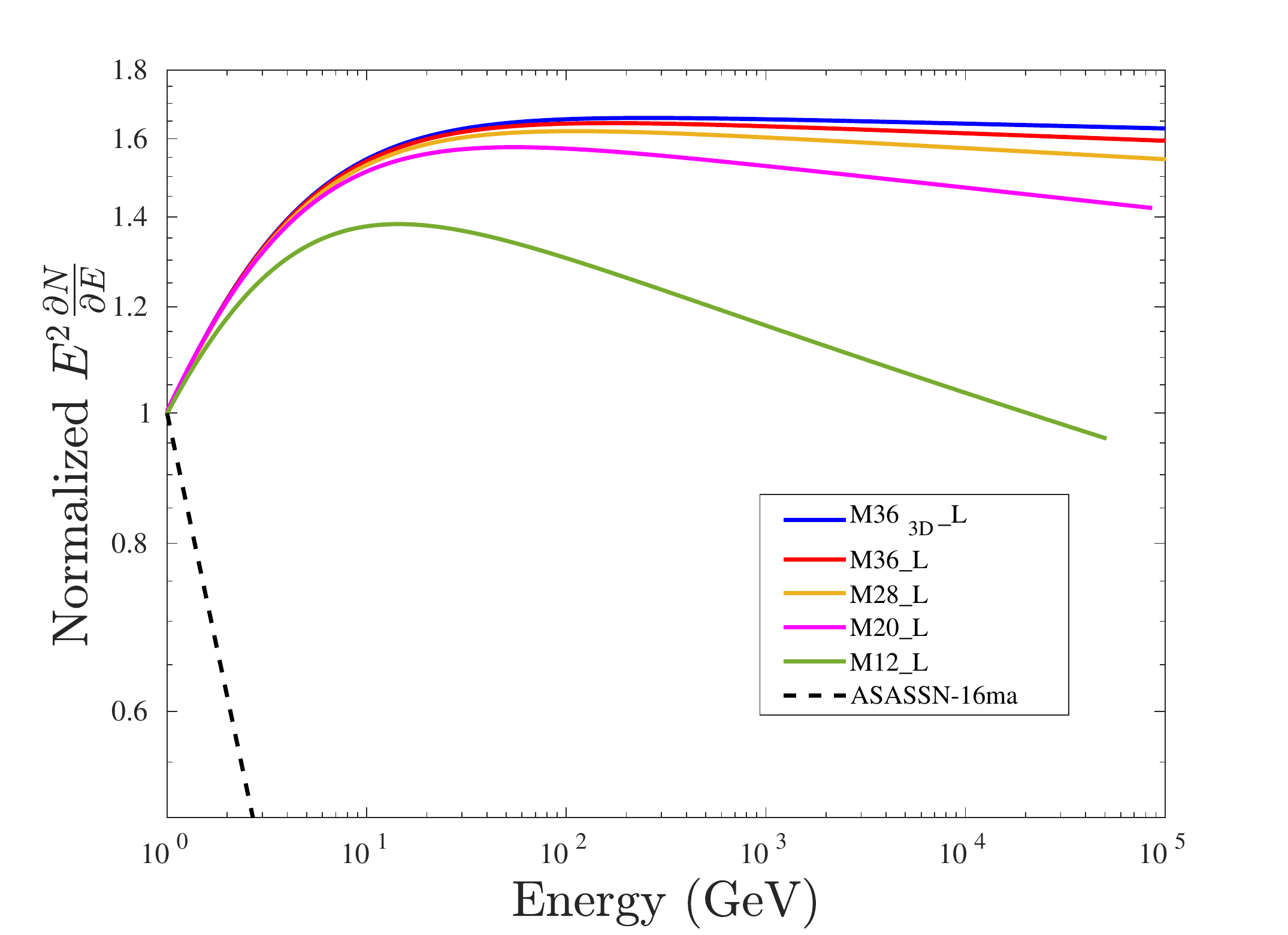}
\caption{Energy spectrum of non-thermal ions accelerated at the outer shock fronts, calculated by applying equation (\ref{eq:r}) locally to each patch of the shock interface, for our ``Lagrangian" runs with different Mach numbers.  For comparison with a dashed line is the much steeper slope inferred by modeling the gamma-ray spectrum of ASASSN-16ma \citep{Li+17}.  As we discuss further in the text, this discrepancy could be due to the preferential escape of high energy cosmic rays (before producing gamma-rays), or due to an additional source of particle acceleration with a steeper index in the turbulent post-shock region.}
\label{fig:spectrum}
\end{figure}

Figure \ref{fig:spectrum} shows our calculation of the average non-thermal spectrum for each of our ``Lagrangian" runs.  These are obtained by combining the shock luminosity-weighted Mach number and inclination distribution (Fig.~\ref{fig:Mach}) with the results of equations (\ref{eq:dNdE}, \ref{eq:r}) for $q(\mathcal{M})$, again only counting regions of the shock which obey the $\theta_{\rm B} \lesssim \theta_{\rm max}$ condition (we also give in Table \ref{table:results} the shock luminosity-weighted value of the ion momentum index $q$).  

Despite the reduction in Mach number due to the irregular shape of the shock front, we find that the energy spectrum of relativistic particles for our high $\mathcal{M}$ simulations is very close to the $dN/dE \propto E^{-\beta}$ with $\beta \simeq 2$ expected in the $\mathcal{M} \gg 1$ limit.  This behavior is inconsistent with the steeper value $\beta \approx 2.7$ of the ion spectrum inferred by modeling the gamma-ray emission of ASASSN-16ma \citep{Li+17}.  Part of this discrepancy could be due to the energy-dependent escape of cosmic ray ions from the shock front, such that higher energy particles with large Larmor radii have a greater chance of escaping the surrounding medium without producing gamma-rays (see below).  

Alternatively, we may be missing an additional source of particle acceleration with a softer spectrum than the outer shock, such as that due to lower Mach number shocks or turbulence occurring downstream.  If we compare the kinetic power dissipated by the external shocks as measured from the simulation to the total upstream kinetic energy flux, we find that only $\approx 60-80\%$ is accounted for, as tabulated in the final column of Table \ref{table:results}.  This suggests that a sizable fraction of the free kinetic energy of the flow is dissipated in the chaotic downstream, through oblique shocks or turbulence.  We speculate that this could contribute a softer non-thermal ion spectrum than that accelerated by the relatively strong external shocks.

We perform an additional check before proceeding.  We have calculated the properties of shock-accelerated ions by summing the contributions separately from each position along the shock front, as if the acceleration process acts independently at each local position.  This is a reasonable approximation if the ions ``see" an approximately infinite planar shock during the acceleration process.  This in turn requires that the lengthscale over which cosmic rays diffuse upstream in the process of being accelerated, $\lambda_{\rm diff} \approx D/v_{\rm sh}$, be small compared to the length-scale over which the shock or post-shock turbulent region in changing, where here $D = r_{\rm g}c/3$ is the diffusion coefficient (in the limit of Bohm diffusion; \citealt{Caprioli&Spitkovsky14b}) and $r_{\rm g} = E/eB$ is the ion gyro radius, where $e$ is the electron charge.  When thermal instabilities and the NTSI is active, the lengthscale for changes in the shock front or post-shock structure range from the radiative cooling length $\mathcal{L}_{\rm cool}$ (eq.~\ref{eq:Lcool}) to the coherence size of the largest corrugations $\lambda_{\rm max} = v_{\rm sh}t_{\rm exp}/\mathcal{M}$ (Fig.~\ref{fig:hierarchy}).  Considering the cooling length as the relevant one, we find
\begin{eqnarray}
&&\frac{\lambda_{\rm diff}}{\mathcal{L}_{\rm cool}} = 0.3\left(\frac{\epsilon_{B}}{0.01}\right)^{-1/2}\left(\frac{E}{100\,{\rm GeV}}\right)\times \nonumber \\
 &&\left(\frac{n}{10^{10}\,{\rm cm^{-3}}}\right)^{1/2}\left(\frac{v_{\rm sh}}{10^{3}\,{\rm km\,s^{-1}}}\right)^{-7},
\label{eq:rgratio}
\end{eqnarray}
where we have estimated the magnetic field strength near the shock as $B \simeq (6\pi \epsilon_{B}\mu m_p n v_{\rm sh}^{2})^{1/2}$ and $\epsilon_{B} \ll 1$ is the fraction of the kinetic power of the shock placed into magnetic field energy.  

Equation (\ref{eq:rgratio}) shows that, because the energy of ions at the injection scale is very low $E \sim few \times kT_{\rm sh}$, the sub-structure of the shock introduced by thermal instabilities and the NTSI should not affect the {\it injection} of ions and thus the acceleration efficiency.  However, higher energy particles $E \lesssim 100-1000$ GeV which produce the gamma-rays detectable by {\it Fermi} LAT can in principle have $\lambda_{\rm diff} \gtrsim \mathcal{L}_{\rm cool}$ for shock velocities and densities relevant to classical novae and other transients.  The preferential escape of higher energy particles during the acceleration process could therefore play a role in the steeper than expected ion spectrum inferred for classical novae such as ASSASN-16ma (Fig.~\ref{fig:spectrum}).  

\section{Discussion and Conclusions}
\label{sec:discussion}


We have presented the results of two and three-dimensional moving-mesh hydrodynamical simulations of the radiative shocks created by dual equal-density colliding flows.  Our goal was to assess the impact of the thin-shell instability on the temperature distribution of the thermal radiation and the efficiency of non-thermal ion acceleration relevant to the non-thermal X-ray and gamma-ray radiation.  

Our main findings are summarized as follows:
\begin{itemize}

\item{Radiative shocks play an important role in a variety of astrophysical transients, such as Type IIn supernovae, classical novae, and stellar mergers.  When the photon diffusion time away from the shock is much less than the expansion time (Fig.~\ref{fig:schematic}), the shock dynamics are well-approximated assuming optically-thin radiative cooling.
}

\item{High Mach number radiative shocks are susceptible to thermal instabilities \citep{Chevalier&Immamura82} and the NTSI \citep{Vishniac94}, the latter of which in its saturated non-linear state creates a highly corrugated shock front (a ``mountain range" in 3D) and a complex turbulent downstream characterized by cold filaments immersed in a hot volume-filling gas (Figs.~\ref{fig:walkthrough},\ref{fig:walkthrough3D},\ref{fig:tempcompare}).  

The radial extent of the cold filamentary region is enhanced as compared to the naive one-dimensional laminar compression picture, but it remains a factor of at least $\sim 1/\mathcal{M}$ thinner than it would otherwise be for an adiabatic shock (Fig.~\ref{fig:hierarchy}).  This difference between the naive one-dimensional laminar flow expectation, and the true multi-dimensional post-shock structure, could affect estimates of the clumping of the cool gas and the (eventual) process of dust formation \citep{Derdzinski+17}.  Application of our results to these issues will be the aim of future work.}
\item{Properly capturing the multi-phase structure of the post-shock region requires suppressing artificial numerical conduction from the hot to the cold regions \citep{Parkin&Pittard10}, which is challenging because of the large density contrast involved and the high cooling rate of the cold gas.  We mitigate this effect in our simulations by employing a new ``Lagrangian" numerical technique that eliminates the advective term in the numerical scheme at the expense of only moderate loss of accuracy (Appendix \ref{sec:appendix}).  The impact of this technique on preserving the high temperature phase of the post-shock region is striking (Fig.~\ref{fig:tempcompare}).}
\item{The average temperature of the radiated thermal emission (Figs.~\ref{fig:NTSI_T_M36},\ref{fig:NTSI_T_Mall}), and the fraction of the shock's thermal luminosity radiated at the maximum temperature $\approx T_{\rm sh}$ (Fig.~\ref{fig:T_bar}) is strongly suppressed compared to the naive expectation from a 1D planar compression analysis. The degree of suppression is well-approximated by the expression $L(>T_{\rm sh}/3)/L_{\rm tot} \sim 4.5/\mathcal{M}^{4/3}$ for $4 \lesssim \mathcal{M} \lesssim 36$ (Fig.~\ref{fig:T_reduction}).  Our findings broadly agree with those of previous work (e.g.~\citealt{Kee+14}), but we illustrate the Mach number and 2D versus 3D dependence of this effect here for the first time.  

We find that the temperature suppression only partially results from the lower Mach numbers of the oblique shocks created by the NTSI.  A more important effect appears to be the presence of weak shocks behind the main outer shocks, which are being driven into under-pressurized cold filaments (presumably created by thermal instability; Figs.~\ref{fig:P-T},\ref{fig:zoom}), robbing the hot medium of its thermal energy before it can be radiated.  The suppression appears to be greater in 3D than in 2D, a result which may possibly be attributed to the larger surface area of the cold gas exposed to the hot medium in the 3D case (eq.~\ref{eq:Lcratio}).}

\item{The direction of the upstream magnetic field in the dynamically expanding outflows of astrophysical transients is expected to lie close to the plane of the shock, resulting in the nearly complete suppression of the diffusive shock acceleration of relativistic ions \citep{Caprioli&Spitkovsky14}.  However, the NTSI results in local regions where the obliquity angle is sufficiently large $\theta_{\rm B} \gtrsim 40-55^{\circ}$ to allow ion acceleration (Figs.~\ref{fig:Mach},\ref{fig:schematicshock}).  Averaged over the shock front, the acceleration efficiency is typically $\epsilon_{\rm nt} \sim 1\%$ (Fig.~\ref{fig:e_nt}), depending only weakly on the Mach number.  Though well below the $\sim 10\%$ maximum predicted for quasi-parallel shocks, this acceleration efficiency agrees remarkably well with those inferred from multi-wavelength observations of gamma-ray novae \citep{Metzger+15,Li+17}.  

On the other hand, the ion particle energy spectrum we calculate using standard diffusive shock acceleration theory is too flat compared to observations, e.g. $dN/dE \propto E^{-\beta}$ with $\beta \approx 2$ (Fig.~\ref{fig:spectrum}).  The preferential escape of higher energy ion from the shock region (before they produce gamma-rays), or an additional component of ion acceleration with a softer spectrum from e.g.~weaker shocks within the turbulent region between the outer shocks, could contribute to the steeper spectrum $\beta \approx 2.7$ inferred from the gamma-ray data.  

Our results do not address the implications of radiative shocks for {\it electron} acceleration, for which current PIC simulations suggest may be relatively independent of the upstream magnetic field geometry (e.g.~\citealt{Park+15}).  Indeed, efficient electron acceleration is inferred indirectly in classical novae from radio synchrotron emission at later times than the observed gamma-ray emission (e.g.~\citealt{Chomiuk+14,Weston+16}), but when the forward shock may still be in the radiative regime \citep{Metzger+14,Vlasov+16}.  }       
\end{itemize}

Our results for suppressed X-ray emission have implications for shock-powered astrophysical transients.  As one example, many classical novae are accompanied by thermal X-ray emission of luminosity $L_{\rm X} \sim 10^{33}-10^{35}$ erg s$^{-1}$ and temperatures $\gtrsim 1$ keV (e.g.~\citealt{Sokoloski+06,Schwarz+11,Chomiuk+14}).  The fact that these luminosities are much lower than the power of the shocks $\gtrsim 10^{37}$ erg s$^{-1}$ required to explain the GeV gamma-ray emission has generally led to the conclusion that the shocks are giving rise to these phenomenon are distinct (e.g.~\citealt{Vlasov+16}).  However, from eq.~(\ref{eq:Kee}) we conclude that for Mach numbers $\mathcal{M} \sim 30-200$ of relevance, the X-ray emission will be suppressed by a factor of $\sim 20-200$.  If the observed X-rays arise from the same radiative shocks responsible for the gamma-rays, then the true shock power could be several orders of magnitude higher than would be naively guessed from the measured X-ray luminosity.

Beyond their application to transients, radiative shocks can play a role in a variety of other astrophysical systems, including colliding wind binaries (e.g.~\citealt{Stevens+92,Lamberts+11,Hendrix+16}), old supernova remnants \citep{Raymond+97,Blondin+98}, young stellar object outflows (e.g.~\citealt{Hansen+17}), and cloud-cloud collisions in AGN \citep{Daltabuit&Cox72}.  Our results demonstrate that expectations for the radiation and geometric structure of these phenomenon could require substantial revision when generalizing from expectations based on one-dimensional simulations.

We conclude by addressing some limitations of our work.  First, though we believe our novel ``Lagrangian" technique represents an important advance in eliminating artificial conduction in the post-shock region, we have not demonstrated complete numerical convergence, as resolving the dense cold filamentary structures remains a challenge.  Furthermore, the actual physical effect of electron conduction between the hot and cold phases could be relevant in some cases (eq.~\ref{eq:conduction}), particularly for lower velocity shocks (though magnetic fields could substantially suppress conduction).  

Another limitation of our work is that we do not include pressure support in the post-shock gas from magnetic fields or non-thermal particles (though the cooling time of even relativistic ions may be short compared to the dynamical time in many cases of relevance; \citealt{Vurm&Metzger18}), which could impose a maximum compression ratio lower than that introduced by our temperature floor.  Magnetic pressure, by reducing the density contrast between the cooling layers and the inflowing medium (for fields parallel to the shock normal) or by resisting shear (if the field is in the same plane of the shock), can act to decrease the effectiveness of the NTSI  (\citealt{Ramachandran&Smith05,Heitsch+07}).  On the other hand, ambipolar diffusion or other forms of reconnection may act to eliminate the magnetic field, and its effects on shear may be less important if its coherence length-scale is much smaller than other macroscopic scales in the problem.

\section*{Acknowledgements}
ES and BDM are supported through the NSF grant AST-1615084, NASA Fermi Guest Investigator Program grants NNX16AR73G and 80NSSC17K0501; and through Hubble Space Telescope Guest Investigator Program grant HST-AR-15041.001-A.  We thank Damiano Caprioli, John Raymond, and Lorenzo Sironi for helpful comments and conversations.  We acknowledge computing resources from Columbia University's Shared Research Computing Facility project, which is supported by NIH Research Facility Improvement Grant 1G20RR030893-01, and associated funds from the New York State Empire State Development, Division of Science Technology and Innovation (NYSTAR) Contract C090171, both awarded April 15, 2010.

\bibliographystyle{mnras}
\bibliography{Nova_bib}

\appendix
\section{Lagrangian hydrodynamics on Voronoi moving mesh}
\label{sec:appendix}
\subsection{New ``Lagrangian" Scheme}
While Voronoi based moving mesh codes greatly reduce the advection terms in the Euler equations, they are still present to some extent, resulting in a finite amount of mass transfer between cells.  We provide below an adaption of this scheme which, at the expense of some loss of accuracy, is fully Lagrangian. 

We start by calculating the wave speed of the discontinuity, and solve the Riemann problem at that reference frame. This naturally gives a flux that has no advection terms. However, since the interface between cells moves at a different velocity, an error occurs. In order to minimize this error, we fix the energy flux to compensate for the PdV work that was actually done. In practice, this means that the energy flux is set to be $P^*v_{\rm intr}$ instead of $P^*v^*$, where $P^*$ and $v^*$ are the pressure and velocity at the discontinuity and $v_{\rm intr}$ is the velocity of the interface between cells. 

Since for smooth flows the difference between $v_{\rm intr}$ and $v^*$ should be inversely proportional to the resolution, we expect a first order convergence, while still retaining the conservative character of the scheme.

There are two caveats that need to be addressed.
Firstly, a scenario in which $|v_{\rm cell}|>|v_{\rm intr}|$ can occur when there is a shock wave, where $v_{\rm cell}$ is the velocity of the downstream cell. Since this leads to an unphysical cooling of the shocked cell, we set the energy flux to be $P^*\textrm{max}(v_{\rm intr},v_{\rm cell})$ whenever a shock wave occurs. 
Secondly, it is possible for the sign of $v_{\rm intr}$ and $v^*$ to be opposite. This can give rise to negative thermal energy in the upstream cell. In this case we set the energy flux to be $\textrm{min}(P^*,P_{\rm up})v_{\rm intr}$, where $P_{\rm up}$ is the pressure of the upstream cell.

Since this scheme is handled on an interface basis, in this paper we limit its use to be only on interfaces which have a temperature ratio larger than 3 between the left and the right states. However, for the examples shown below, we employ this scheme on all the interfaces.
\subsection{Lagrangian Mesh Steering}
In order for the error from the Lagrangian scheme be minimized, the Riemann problems should be solved at interfaces moving as closely as possible to the contact speed. We propose the following scheme of setting the velocities of the mesh generating points
\begin{itemize}
\item Calculate the contact speed on the j-th interface, $\vec{w}_j$.
\item For each Voronoi cell, we calculate its predicted new Center of Mass
\begin{eqnarray}
\vec{\textrm{CM}}_{\rm new} &=& \left[V\cdot\vec{\textrm{CM}}_{\rm old}+\sum A_j|\vec{w}_j|dt\left(\vec{F}_j + 0.5\vec{w}_j dt\right)\right]\nonumber \\
&/&\left(
V+\sum A_j{\rm sgn}(\vec{w}_j) |\vec{w}_j|dt\right),
\end{eqnarray}
where $V$ is the volume of the cell, $\vec{F}_j$ is the centroid of the j-th interface, $A_j$ is the area of the j-th interface, $dt$ is the time step and the summation is done over all of the Voronoi cell's interfaces.
\item Set the velocity of the mesh generating point to be
\begin{equation}
\vec{v}_{\rm mesh} = \left(\vec{\textrm{CM}}_{\rm new}-\vec{\textrm{CM}}_{\rm old}\right)/dt.
\end{equation}
\end{itemize}
Additionally, for each cell we calculate its deviation from Lagrangian motion, i.e. $\sum A_j\left(\vec{w}_j-\vec{v}_{{\rm intr},j}\right)dt$, and at the following time step add an additional velocity to compensate for this deviation that is limited in its magnitude to be a fraction of the mesh generating point's velocity. 
\subsection{Convergence for the Yee Vortex}
We benchmark the performance of our new scheme for smooth flows by running the Yee vortex test problem on a randomly distributed mesh.  This problem describes the evolution of isentropic vortex that is pressure supported.  The set up is identical to that described in \cite{Steinberg+16} and is run until time $t=10$.
We calculate the $L_1$ error norm of the density and the angular momentum, defined as $L_1=\frac{1}{V}\int|\phi(x)-\phi(x)_Y|dV$, where $\phi(x)_Y$ is the analytical time independent variable and $\phi(x)$ is the computed variable from the simulation.

\begin{figure}
\centering
\includegraphics[width=0.95\linewidth]{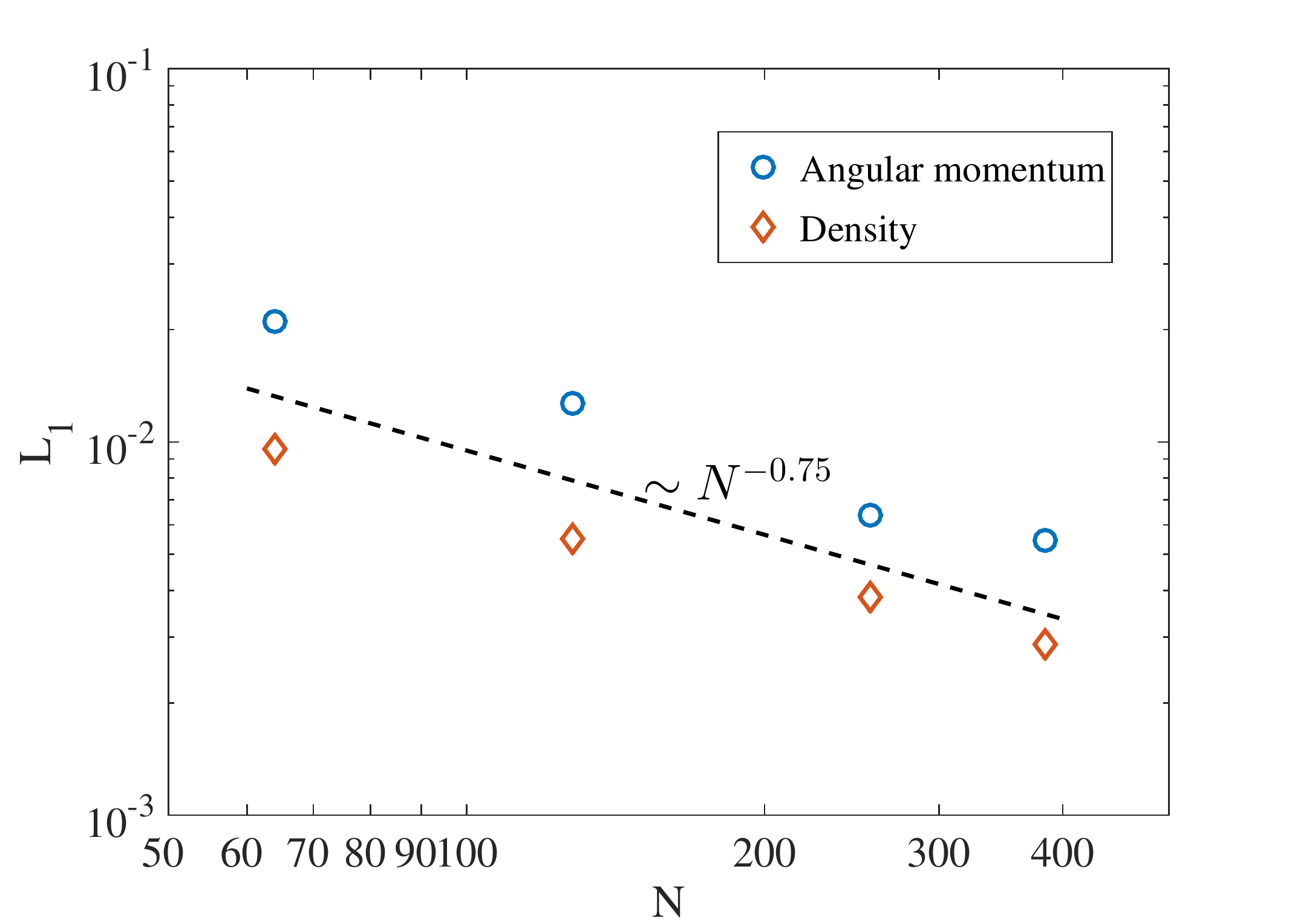}
\caption{The $L_1$ error norm for the density and angular momentum as a function of numerical resolution, for the implementation of our ``Lagrangian" scheme to the Yee vortex test problem.}
\label{fig:Yee}
\end{figure}

Figure \ref{fig:Yee} shows the $L_1$ error norms, demonstrating convergence of both the density and angular momentum, although at a somewhat lower rate of $-0.75$ compared to the expected first order convergence.  This is likely due to the fact that we add an additional small non-Lagrangian velocity to the mesh generating points in order to keep the Voronoi cells `round'.
\subsection{Kelvin-Helmholtz}
\begin{figure}
\centering
\includegraphics[width=0.95\linewidth]{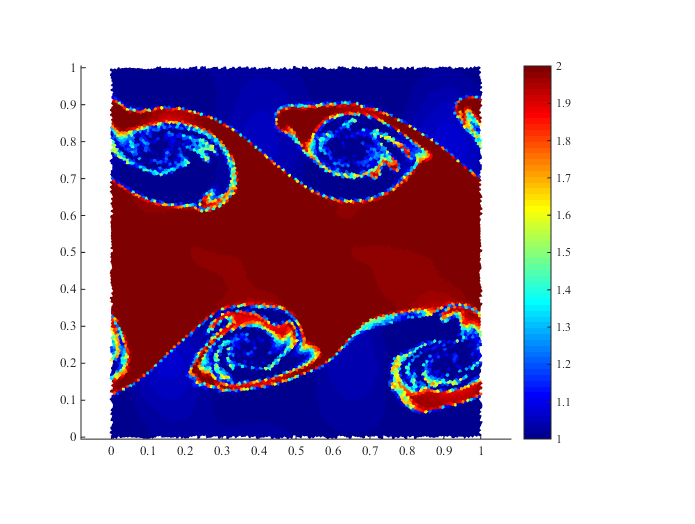}

\includegraphics[width=0.95\linewidth]{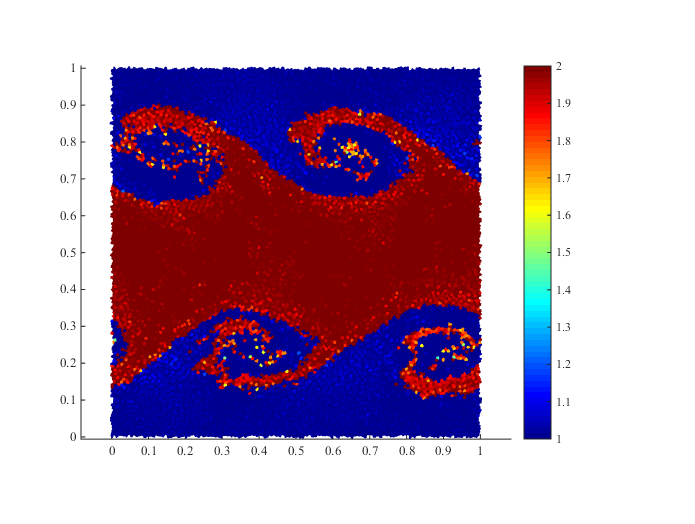}

\caption{Kelvin-Helmholtz test problem, comparing our ``semi-Lagrangian" (top panel) and ``Lagrangian" (bottom panel) schemes.  The color level denotes density.}
\label{fig:KH}
\end{figure}

To test our new scheme's performance for discontinuities, we redo the Kelvin-Helmholtz test problem as presented in \cite{Yalinewich+15}.  We run the problem until time $t=1.5$ using both the ``Lagrangian'' scheme as well as the normal ``semi-Lagrangian'' scheme, both with a resolution of $128^2$.
Figures \ref{fig:KH} shows the density at the completion of the run.  While the ``semi-Lagrangian'' method produces a smoother result than our new ``Lagrangian'' method, the discontinuity is nevertheless perfectly preserved in the ``Lagrangian'' method and the system maintains the same dynamical evolution.

\subsection{Sedov Blast Wave}

\begin{figure}
\centering
\includegraphics[width=0.95\linewidth]{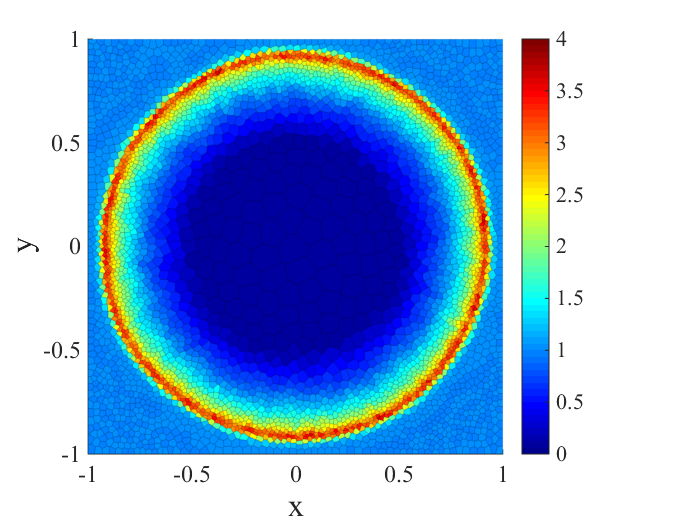}

\includegraphics[width=0.95\linewidth]{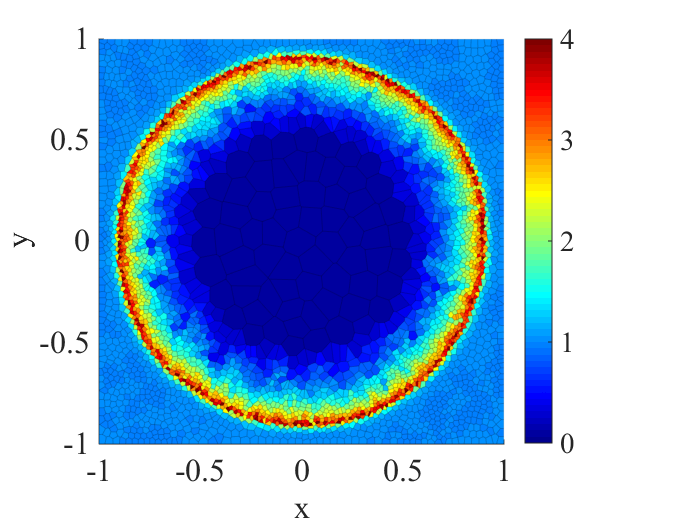}

\caption{Sedov blastwave test problem, comparing our ``semi-Lagrangian" (top panel) and ``Lagrangian" (bottom panel) schemes.  The color level denotes density.}
\label{fig:ST}
\end{figure}

\begin{figure}
\centering
\includegraphics[width=0.95\linewidth]{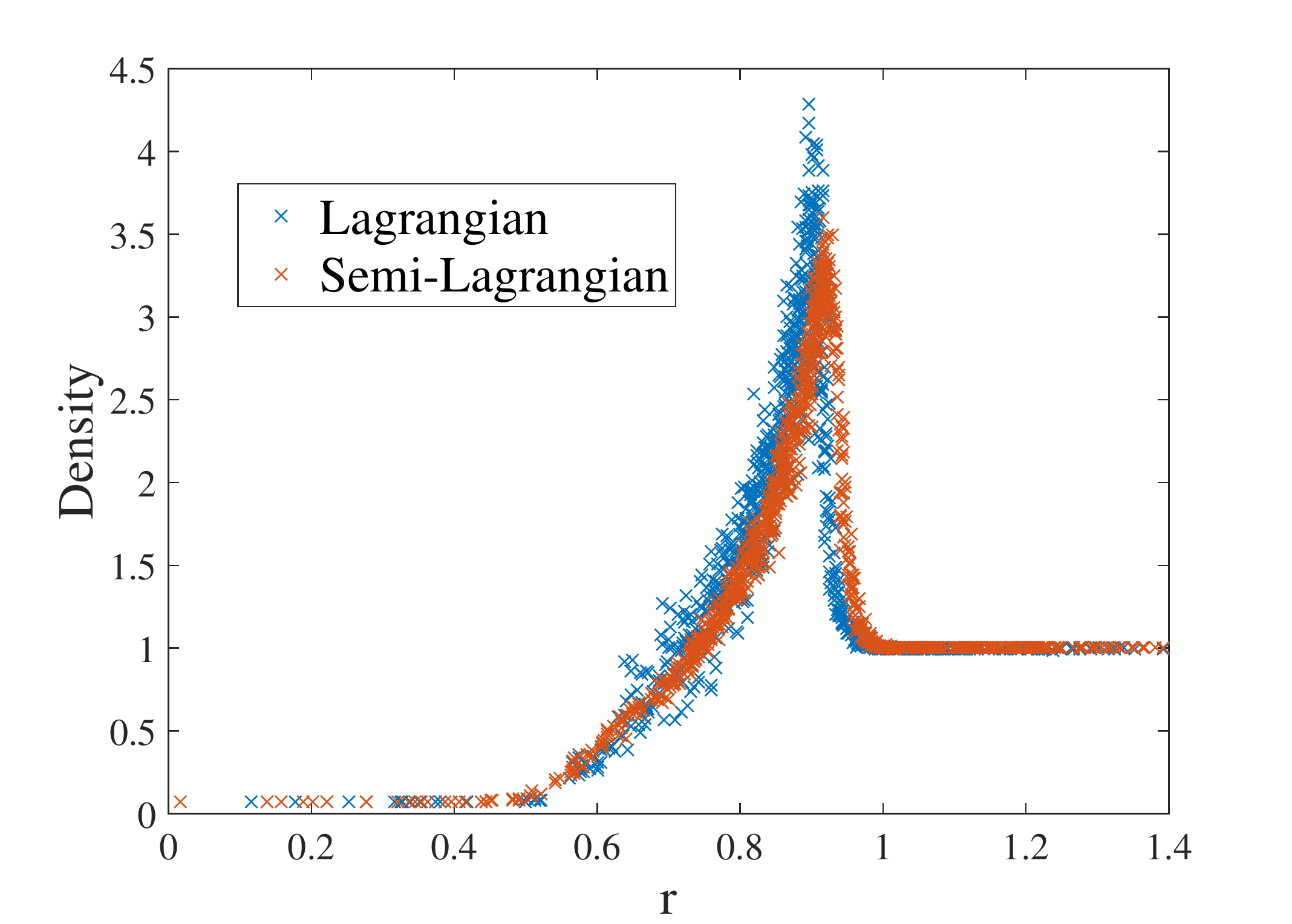}
\caption{Snapshot of the radial density profiles of the Sedov blastwave problem, comparing the results of the semi-Lagrangian and Lagrangian schemes.}
\label{fig:sedov_compare}
\end{figure}

The Sedov blast wave problem provides an ideal way to benchmark how our new scheme handles strong shocks. We set up an initial grid with 2500 randomly distributed mesh generating points with an initial uniform density and a hotspot at the center. Figure \ref{fig:ST} shows density snapshots following some period of evolution, comparing the semi-Lagrangian and the Lagrangian schemes.   While the solution for the semi-Lagrangian scheme is smoother, the Lagrangian scheme does a good job of catching the overall morphology of the blast wave and the shock location. This is shown further in Figure \ref{fig:sedov_compare}, which compares the radial density profiles between the  schemes.  The peak density achieved in Lagrangian case is about 25\% higher than in the Semi-Lagrangian case, but the overall shape is similar.  

\bsp	
\label{lastpage}
\end{document}